\documentclass[xcolor={table}]{tufte-handout}

\usepackage{graphicx}
\usepackage{tabularx}
\usepackage{booktabs}
\usepackage{makecell}
\usepackage{amsmath, amsfonts, amssymb, amsthm}
\usepackage{units}
\usepackage{booktabs}
\usepackage{comment}
\usepackage{hyperref}
\usepackage[normalem]{ulem}
\definecolor{myblue}{HTML}{7090CF}
\usepackage{float}
\usepackage{changepage}
\usepackage{tocloft}

\setkeys{Gin}{width=\linewidth, totalheight=\textheight, keepaspectratio}
\graphicspath{{Figures/}{./}}
\definecolor{mybg}{rgb}{0.99, 0.985, 0.98}

\newcommand{\mycaption}[1]{%
    \vspace{0.5em}
    \noindent{\textcolor{gray}{#1}} 
    \vspace{0.5em}
}

\title{\huge{\mbox{\bf Keep the Future Human:} Why and How We Should Close the Gates to AGI and Superintelligence, and What We Should Build Instead}\vspace{0.2cm}}

\author{\fontsize{14pt}{18pt}\selectfont Anthony Aguirre}
\date{March 5, 2025}

\begin{document}

\maketitle

    \setcounter{tocdepth}{1}  
    \tableofcontents

    \newpage
\section{Executive summary}

Dramatic advances in artificial intelligence over the past decade (for narrow-purpose AI) and the last several years (for general-purpose AI) have transformed AI from a niche academic field to the core business strategy of many of the world’s largest companies, with hundreds of billions of dollars in annual investment in the techniques and technologies for advancing AI’s capabilities.

We now come to a critical juncture. As the capabilities of new AI systems begin to match and exceed those of humans across many cognitive domains, humanity must decide: how far do we go, and in what direction? 

AI, like every technology, started with the goal of improving things for its creator. But our current trajectory, and implicit choice, is an unchecked race toward ever-more powerful systems, driven by economic incentives of a few huge technology companies seeking to automate large swathes of current economic activity and human labor. If this race continues much longer, there is an inevitable winner: AI itself -- a faster, smarter, cheaper alternative to people in our economy, our thinking, our decisions, and eventually in control of our civilization.

But we can make another choice: via our governments, we can take control of the AI development process to impose clear limits, lines we won’t cross, and things we simply won’t do -- as we have for nuclear technologies, weapons of mass destruction, space weapons, environmentally destructive processes, the bioengineering of humans, and eugenics. Most importantly, we can ensure that AI remains a tool to empower humans, rather than a new species that replaces and eventually supplants us.

This essay argues that we should {\em keep the future human} by closing the "gates" to smarter-than-human, autonomous, general-purpose AI -- sometimes called “AGI” -- and especially to the highly-superhuman version sometimes called “superintelligence.” Instead, we should focus on powerful, trustworthy AI tools that can empower individuals and transformatively improve human societies' abilities to do what they do best. The structure of this argument follows in brief.

\subsection*{AI is different}

AI systems are fundamentally different from other technologies. While traditional software follows precise instructions, AI systems learn how to achieve goals without being explicitly told how. This makes them powerful: if we can cleanly define the goal or a metric of success, in most cases an AI system can learn to achieve it. But it also makes them inherently unpredictable: we cannot reliably determine what actions they will take to achieve their objectives. 

They are also largely unexplainable: although they are partly code, they are mostly an enormous set of inscrutable numbers -- neural network "weights" -- that cannot be parsed; we are not much better at understanding their inner workings than at discerning thoughts by peering inside a biological brain. 

This core mode of training digital neural networks is rapidly increasing in complexity. The most powerful AI systems are created through massive computational experiments, using specialized hardware to train neural networks on enormous datasets, which are then augmented with software tools and superstructure. 

This has led to the creation of very powerful tools for creating and processing text and images, performing mathematical and scientific reasoning, aggregating information, and interactively querying a vast store of human knowledge.

Unfortunately, while development of more powerful, more trustworthy technological tools is what we {\em should} do, and what nearly everybody wants and says they want, it is not the trajectory we are actually on.

\subsection*{AGI and superintelligence}

Since the dawn of the field, AI research has instead focused on a different goal: Artificial General Intelligence. This focus has now become the focus of the titanic companies leading AI development.

What is AGI? It is often vaguely defined as "human-level AI," but this is problematic: which humans, and at which capabilities is it human level? And what about the super-human capabilities it already has? A more useful way to understand AGI is through the intersection of three key properties: high {\bf A}utonomy (independence of action), high {\bf G}enerality (broad scope and adaptability), and high {\bf I}ntelligence (competence at cognitive tasks). Current AI systems may be highly capable but narrow, or general but requiring constant human oversight, or autonomous but limited in scope. 

Full A-G-I would combine all three properties at levels matching or exceeding top human capability. Critically, it is this combination that makes humans so effective and so different from current software; it is also what would enable people to be wholesale replaced by digital systems.

While human intelligence is special, it is by no means a limit. Artificial "superintelligent" systems could operate hundreds of times faster, parse vastly more data and hold enormous quantities "in mind" at once, and form aggregates that are much larger and more effective than collections of humans. They could supplant not individuals but companies, nations, or our civilization as a whole.

\subsection*{We are at the threshold}

There is a strong scientific consensus that AGI is {\em possible.} AI already surpasses human performance in many general tests of intellectual capability, including recently high-level reasoning and problem solving. Lagging capabilities -- such as continual learning, planning, self-awareness, and originality -- all exist at some level in present AI systems, and known techniques exist that are likely to improve all of them.

While until a few years ago many researchers saw AGI as decades away, currently evidence for short timelines to AGI is strong:

\begin{itemize}

\item Empirically verified "scaling laws" connect computational input to AI capability, and corporations are on-track to scale computational input by orders of magnitude over the coming several years. The human and fiscal resources dedicated to AI advancement now equal those of a dozen Manhattan Projects and several Apollo Projects.

\item AI corporations and their leaders publicly and privately believe that AGI (by some definition) is achievable within a few years. These companies have information the public does not, including some having the next generation of AI systems in-hand.

\item Expert predictors with proven track-records assign 25\% probability to AGI (by some definition) arriving within 1-2 years, and 50\% for 2-5 years (see Metaculus predictions for \href{https://www.metaculus.com/questions/3479/date-weakly-general-ai-is-publicly-known/}{'weak'} and \href{https://www.metaculus.com/questions/5121/date-of-artificial-general-intelligence/}{'full'} AGI).

\item Autonomy (including long-range flexible planning) lags in AI systems, but major companies are now focusing their vast resources on developing autonomous AI systems and have informally named 2025 the \href{https://techinformed.com/2025-informed-the-year-of-agentic-ai/}{"year of the agent."}

\item AI is contributing more and more to its own improvement. Once AI systems are as competent as human AI researchers at doing AI research, a critical threshold for fast progress to much more powerful AI systems will be hit and likely lead to a runaway in AI capability. (Arguably, that runaway has already begun.)

\end{itemize}

The idea that smarter-than-human AGI is decades away or more is simply no longer tenable to the vast majority of experts in the field. Disagreements now are about how many months or years it will take if we stay on this course. The core question we face is: should we?

\subsection*{What is driving the race to AGI}

\vspace{-0.2cm}
The race toward AGI is being driven by multiple forces, each making the situation more dangerous. Major technology companies see AGI as the ultimate automation technology -- not just augmenting human workers but replacing them largely or entirely. For companies, the prize is enormous: the opportunity to capture a significant fraction of the world's \$100 trillion annual economic output by automating away human labor costs.

Nations feel compelled to join this race, publicly citing economic and scientific leadership, but privately viewing AGI as a potential revolution in military affairs comparable to nuclear weapons. Fear that rivals might gain a decisive strategic advantage creates a classic arms race dynamic.

Those pursuing superintelligence often cite grand visions: curing all diseases, reversing aging, achieving breakthroughs in energy and space travel, or creating superhuman planning capabilities. 

Less charitably, what drives the race is power. Each participant -- whether company or country -- believes that intelligence equals power, and that they will be the best steward of that power. 

I argue that these motivations are real but fundamentally misguided: AGI will {\em absorb} and {\em seek} power rather than grant it; AI-created technologies will {\em also} be strongly double-edged, and where beneficial can be created with AI tools and without AGI; and even insofar as AGI and its outputs remain under control, these racing dynamics -- both corporate and geopolitical -- make large-scale risks to our society nearly inevitable unless decisively interrupted.

\vspace{-0.25cm}
\subsection*{AGI and superintelligence pose a dramatic threat to civilization}

Despite their allure, AGI and superintelligence pose dramatic threats to civilization through multiple reinforcing pathways:

{\em Power concentration:} superhuman AI could disempower the vast majority of humanity by absorbing huge swathes of social and economic activity into AI systems run by a handful of giant companies (which may in turn either be taken over by, or effectively take over, governments.) 

{\em Massive disruption:} bulk automation of most cognitive-based jobs, replacement of our current epistemic systems, and rollout of vast numbers of active nonhuman agents would upend most of our current civilizational systems in a relatively short period of time.

{\em Catastrophes:} by proliferating the ability -- potentially above human level -- to create new military and destructive technologies and decoupling it from the social and legal systems grounding responsibility, physical catastrophes from weapons of mass destruction become dramatically more likely.

{\em Geopolitics and war:} major world powers will not sit idly by if they feel that a technology that could supply a "decisive strategic advantage" is being developed by their adversaries.

{\em Runaway and loss of control:} Unless it is specifically prevented, superhuman AI will have every incentive to further improve itself and could far outstrip humans in speed, data processing, and sophistication of thinking. There is no meaningful way in which we can be in control of such a system. Such AI will not grant power to humans; we will grant power to it, or it will take it.

Many of these risks remain even if the technical "alignment" problem -- ensuring that advanced AI reliably does what humans want it to do -- is solved. AI presents an enormous challenge in how it will be managed, and very many aspects of this management become incredibly difficult or intractable as human intelligence is breached.

Most fundamentally, the type of superhuman general-purpose AI currently being pursued would, by its very nature, have goals, agency, and capabilities exceeding our own. It would be inherently uncontrollable -- how can we control something that we can neither understand nor predict? It would not be a technological tool for human use, but a second species of intelligence on Earth alongside ours. If allowed to progress further, it would constitute not just a second species but a replacement species. 

Perhaps it would treat us well, perhaps not. But the future would belong to it, not us. The human era would be over.

\subsection*{This is not inevitable; humanity can, very concretely, decide not to build our replacement.}

The creation of superhuman AGI is far from inevitable. We can prevent it through a coordinated set of governance measures:

First, we need robust accounting and oversight of AI computation ("compute"), which is a fundamental enabler of, and lever to govern, large-scale AI systems. This in turn requires standardized measurement and reporting of the total compute used in training AI models and running them, and technical methods of tallying, certifying, and verifying computation used.

Second, we should implement hard caps on AI computation, both for training and for operation; these prevent AI both from being too powerful and operating too quickly. These caps can be implemented through both legal requirements and hardware-based security measures built into AI-specialized chips, analogous to security features in modern phones. Because specialized AI hardware is made by only a handful of companies, verification and enforcement are feasible through the existing supply chain.

Third, we need enhanced liability for the most dangerous AI systems. Those developing AI that combines high autonomy, broad generality, and superior intelligence should face strict liability for harms, while safe harbors from this liability would encourage development of more limited and controllable systems.

Fourth, we need tiered regulation based on risk levels. The most capable and dangerous systems would require extensive safety and controllability guarantees before development and deployment, while less powerful or more specialized systems would face proportionate oversight. This regulatory framework should eventually operate at both national and international levels.

This approach -- with detailed specification given in the full document -- is practical: while international coordination will be needed, verification and enforcement can work through the small number of companies controlling the specialized hardware supply chain. It is also flexible: companies can still innovate and profit from AI development, just with clear limits on the most dangerous systems.

Longer-term containment of AI power and risk would require international agreements based on both self- and common-interest, just as controlling nuclear weapon proliferation does now. But we can start immediately with enhanced oversight and liability, while building toward more comprehensive governance.

The key missing ingredient is political and social will to take control of the AI development process. The source of that will, if it comes in time, will be reality itself -- that is, from widespread realization of the real implications of what we are doing.

\subsection*{We can engineer Tool AI to empower humanity}

Rather than pursuing uncontrollable AGI, we can develop powerful "Tool AI" that enhances human capability while remaining under meaningful human control. Tool AI systems can be extremely capable while avoiding the dangerous triple-intersection of high autonomy, broad generality, and superhuman intelligence, as long as we engineer them to be controllable at a level commensurate with their capability. They can also be combined into sophisticated systems that maintain human oversight while delivering transformative benefits.

Tool AI can revolutionize medicine, accelerate scientific discovery, enhance education, and improve democratic processes. When properly governed, it can make human experts and institutions more effective rather than replacing them. While such systems will still be highly disruptive and require careful management, the risks they pose are fundamentally different from AGI: they are risks we can govern, like those of other powerful technologies, not existential threats to human agency and civilization. And crucially, when wisely developed, AI tools can help people govern powerful AI and manage its effects.

This approach requires rethinking both how AI is developed and how its benefits are distributed. New models of public and non-profit AI development, robust regulatory frameworks, and mechanisms to distribute economic benefits more broadly can help ensure AI empowers humanity as a whole rather than concentrating power in a few hands. AI itself can help build better social and governance institutions, enabling new forms of coordination and discourse that strengthen rather than undermine human society. National security establishments can leverage their expertise to make AI tool systems genuinely secure and trustworthy, and a true source of defense as well as national power.

We may eventually choose to develop yet more powerful and more sovereign systems that are less like tools and -- we can hope -- more like wise and powerful benefactors. But we should do so only after we have developed the scientific understanding and governance capacity to do so safely. Such a momentous and irreversible decision should be made deliberately by humanity as a whole, not by default in a race between tech companies and nations.

\subsection*{In human hands}

People want the good that comes from AI: useful tools that empower them, supercharge economic opportunities and growth, and promise breakthroughs in science, technology, and education. Why wouldn’t they? But when asked, overwhelming majorities of the general public \href{https://www.vox.com/future-perfect/2023/8/18/23836362/ai-slow-down-poll-regulation}{want slower and more careful AI development}, and do not want smarter-than-human AI that will replace them in their jobs and elsewhere, fill their culture and information commons with non-human content, concentrate power in a tiny set of corporations, pose extreme large-scale global risks, and eventually threaten to disempower or replace their species. Why would they?

We {\em can} have one without the other. It starts by deciding that our destiny is not in the supposed inevitability of some technology or in the hands of a few CEOs in Silicon Valley, but in the rest of our hands if we take hold of it. Let’s close the Gates, and keep the future human.

\newpage

\section{Introduction}

We may be at the end of the human era. 

Something has begun in the past ten years that is unique in the history of our species. Its consequences will, to a great extent, determine the future of humanity. Starting around 2015, researchers have succeeded in developing {\em narrow} artificial intelligence (AI) -- systems that can win at games like Go, recognize images and speech, and so on, better than any human.\sidenote[][-15mm]{This \href{https://time.com/6300942/ai-progress-charts/}{chart} shows a set of tasks; many similar curves could be added to this graph. This rapid progress in narrow AI has surprised even experts in the field, with benchmarks being surpassed years ahead of predictions.}

This is amazing success, and it is yielding extremely useful systems and products that will empower humanity. But narrow artificial intelligence has never been the true goal of the field. Rather, the aim has been to create {\em general} purpose AI systems, particularly ones often called "artificial general intelligence" (AGI) or "superintelligence" that are simultaneously as good or better than humans across nearly {\em all} tasks, just as AI is now super-human at Go, chess, poker, drone racing, etc. This is the stated goal of many major AI companies.\sidenote[][-15mm]{Deepmind, OpenAI, Anthropic, and X.ai were all founded with the specific goal of developing AGI. For instance, OpenAI's charter explicitly states its goal as developing "artificial general intelligence that benefits all of humanity," while DeepMind's mission is "to solve intelligence, and then use that to solve everything else." Meta, Microsoft, and others are now pursuing substantially similar paths. Meta has said that it \href{https://www.forbes.com/sites/johnkoetsier/2024/01/18/zuckerberg-on-ai-meta-building-agi-for-everyone-and-open-sourcing-it/}{plans to develop AGI and release it openly.}}

{\em These efforts are also succeeding.} General-purpose AI systems like ChatGPT, Gemini, Llama, Grok, Claude, and Deepseek, based on massive computations and mountains of data, have reached parity with typical humans across a wide variety of tasks, and even match human experts in some domains. Now AI engineers at some of the largest technology companies are racing to push these giant experiments in machine intelligence to the next levels, at which they match and then exceed the full range of human capabilities, expertise, and autonomy. 

{\em This is imminent.} Over the last ten years, expert estimates for how long this will take -- if we continue our present course -- have fallen from decades (or centuries) to single-digit years. 

It is also of epochal importance, and transcendent risk. Proponents of AGI see it as a positive transformation that will solve scientific problems, cure disease, develop new technologies, and automate drudgery. And AI could certainly help to achieve all of these things -- indeed it already is. But over the decades, many careful thinkers, from Alan Turing to Stephen Hawking to the present-day Geoffrey Hinton and Yoshua Bengio\sidenote[][-55mm]{Hinton and Bengio are two of the most cited AI researchers, have both won the AI field's Nobel, the Turing Prize, and Hinton has won a Nobel prize (in physics) to boot.} have issued a stark warning: building truly smarter-than-human, general, autonomous AI will at minimum completely and irrevocably upend society, and at maximum result in human extinction.\sidenote[][-100pt]{Building something of this risk, under commercial incentives and near-zero government oversight, is utterly unprecedented. There isn't even controversy about the risk among those building it! The leaders of Deepmind, OpenAI, and Anthropic, among many other experts, have all literally signed a \href{https://www.safe.ai/work/statement-on-ai-risk}{statement} that advanced AI poses an {\em extinction risk to humanity.} The alarm bells could not be ringing any harder, and one can only conclude that those ignoring them simply are not taking AGI and superintelligence seriously. One goal of this essay is to help them understand why they should.}

Superintelligent AI is rapidly approaching on our current path, but is far from inevitable. This essay is an extended argument as to why and how we should \textit{close the Gates} to this approaching inhuman future, and what we should do instead.

\section{Need-to-knows about AI neural networks}

To understand how the consequences of developing more powerful AI will play out, it is essential to internalize some basics. This and the next two sections develop these, covering in turn what modern AI is, how it leverages massive computations, and the senses in which it is rapidly growing in generality and capability.\sidenote[][-20mm]{For a gentle but technical introduction to machine learning and AI, particularly language models, see \href{https://mark-riedl.medium.com/a-very-gentle-introduction-to-large-language-models-without-the-hype-5f67941fa59e}{this site.} For another modern primer on AI extinction risks, see \href{https://www.thecompendium.ai/}{this piece.} For a comprehensive and authoritative scientific analysis of the state of AI safety, see the recent \href{https://arxiv.org/abs/2501.17805}{International
AI Safety Report.}}

There are many ways to define artificial intelligence, but for our purposes the key property of AI is that while a standard computer program is a list of instructions for how to perform a task, an AI system is one that learns from data or experience to perform tasks {\em without being explicitly told how to do so.} 

Almost all salient modern AI is based on neural networks. These are mathematical/computational structures, represented by a very large (billions or trillions) set of numbers ("weights"), that perform a training task well. These weights are crafted (or perhaps "grown" or "found") by iteratively tweaking them so that the neural network improves a numerical score (a.k.a. "loss") defined toward performing well at one or more tasks.\sidenote[][-130pt]{Training typically occurs by looking for a local maximum of the score in a high-dimensional space given by the model weights. By checking how the score changes as weights are tweaked, the training algorithm identifies which tweaks improve score the most, and moves the weights in that direction.} This process is known as {\em training} the neural network.\sidenote[][-19mm]{For example, in an image recognition problem, the neural network would output probabilities for labels for the image. A score would be related to the probability the AI accords to the correct answer. The training procedure would then adjust weights so that next time, the AI would output a higher probability for the correct label for that image. This is then repeated a huge number of times. The same basic procedure is used in training essentially all modern neural networks, albeit with more complex scoring mechanism.}

There are many techniques for doing this training, but those details are much less relevant than the ways in which the scoring is defined, and how those result in different tasks the neural network performs well. A key difference has historically been drawn between "narrow" and "general" AI.

Narrow AI is deliberately trained to do a particular task or small set of tasks (such as recognizing images or playing chess); it requires retraining for new tasks, and has a narrow scope of capability. We have superhuman narrow AI, meaning that for nearly any discrete well-defined task a person can do, we can probably construct a score and then successfully train a narrow AI system to do it better than a human could.

General-purpose AI (GPAI) systems can perform a wide range of tasks, including many they were not explicitly trained for; they can also learn new tasks as part of their operation. Current large "multimodal models"\sidenote[][-130pt]{Most multimodal models use the "transformer" architecture to process and generate multiple types of data (text, images, sound). These can all decomposed into, and then treated on the same footing, as different types of "tokens." Multimodal models are trained first to accurately predict tokens within massive datasets, then refined through reinforcement learning to enhance capabilities and shape behaviors.} like ChatGPT exemplify this: trained on a very large corpus of text and images, they can engage in complex reasoning, write code, analyze images, and assist with a vast array of intellectual tasks. While still quite different from human intelligence in ways we'll see in depth below, their generality has caused a revolution in AI.\sidenote[][-26mm]{That language models are trained to do one thing -- predict words -- has caused some to call them narrow AI. But this is misleading: because predicting text well requires so many different capabilities, this training task leads to a surprisingly general system. Also note that these systems are extensively trained by reinforcement learning, effectively representing thousands of people giving the model a reward signal when it does a good job at any of the many things it does. It then inherits significant generality from the people giving this feedback.}

\subsection*{Unpredictability: a key feature of AI systems}

A key difference between AI systems and conventional software is in predictability. Standard software's output can be unpredictable -- indeed sometimes that's why we write software, to give us results we could not have predicted. But conventional software rarely does anything it was not programmed to do -- its scope and behavior are generally as designed. A top-tier chess program may make moves no human could predict (or else they could beat that chess program!) but it will not generally do anything but play chess.

Like conventional software, narrow AI has predictable scope and behavior but can have unpredictable results. This is really just another way to define narrow AI: as AI that is akin to conventional software in its predictability and range of operation.

General-purpose AI is different: its scope (the domains over which it applies), behavior (the sorts of things it does), and results (its actual outputs) can all be unpredictable.\sidenote{There are multiple ways in which AI is unpredictable. One is that in the general case one cannot predict what an algorithm will do without actually running it; there are \href{https://arxiv.org/abs/1310.3225}{theorems} to this effect. This can be true just because the output of algorithms can be complex.  But it is particularly clear and relevant in the case (such as in chess or Go) where the prediction would imply a capability (beating the AI) the would-be predictor does not have. Second, a given AI system will not always produce the same output even given the same input -- its outputs contain randomness; this also couples with algorithmic unpredictability. Third, unexpected and emergent capabilities can arise from training, meaning even the {\em types} of things an AI system can and will do are unpredictable; This last type is particularly important for safety considerations.} GPT-4 was trained just to generate text accurately, but developed many capabilities its trainers didn't predict or intend. This unpredictability stems from the complexity of training: because the training data contains outputs from many different tasks, the AI must effectively learn to perform these tasks to predict well.

This unpredictability of general AI systems is quite fundamental.  Although in principle it is possible to carefully construct AI systems that have guaranteed limits on their behavior (as mentioned later in the essay), the way AI systems are created now they are unpredictable in practice and even in principle.

\subsection*{Passive AI, agents, autonomous systems, and alignment}

This unpredictability becomes particularly important when we consider how AI 
systems are actually deployed and used to achieve various goals.

Many AI systems are relatively passive in the sense that they primarily provide information, and the user takes actions. Others, commonly termed {\em agents}, take actions themselves, with varying levels of involvement from a user. Those that take actions with relatively less external input or oversight may be termed more {\em autonomous}. This forms a spectrum in terms of independence of action, from passive tools to autonomous agents.\sidenote{See \href{https://arxiv.org/abs/2502.02649}{here} for an in-depth review of what is meant by an "autonomous agent" (along with ethical arguments against building them).}

As for goals of AI systems, these may be directly tied to their training objective (e.g. the goal of "winning" for a Go-playing system is also explicitly what it was trained to do). Or they may not be: ChatGPT's training objective is in part to predict text, in part to be a helpful assistant. But when doing a given task, its goal is supplied to it by the user. Goals may also be created by an AI system itself, only very indirectly related to its training objective.\sidenote{You may sometimes hear "AI can't have its own goals." This is absolute nonsense. It is easy to generate examples where AI has or develops goals that were never given to it and are known only to itself. You don't see this much in current popular multimodal models because it is trained out of them; it could just as easily be trained into them.}

Goals are closely tied to the question of "alignment," that is the question of whether AI systems will {\em do what we want them to do}. This simple question hides an enormous level of subtlety.\sidenote[][1mm]{There's a large literature. On the general problem see Christian's \href{https://www.amazon.com/Alignment-Problem-Machine-Learning-Values/dp/0393635821}{\em The Alignment Problem}, and Russell's \href{https://www.amazon.com/Human-Compatible-Artificial-Intelligence-Problem/dp/0525558616}{\em Human-Compatible}. On a more technical side see e.g. \href{https://arxiv.org/abs/2209.00626}{this paper}.}
For now, note that "we" in this sentence might refer to many different people and groups, leading to different types of alignment. For example, an AI might be highly {\em obedient} (or \href{https://arxiv.org/abs/2003.11157}{"loyal"}) to its user -- here "we" is "each of us." Or it might be more {\em sovereign}, being primarily driven by its own goals and constraints, but 
still acting broadly in the common interest of human wellbeing -- "we" is then "humanity" or "society."  In-between is a spectrum where an AI would be largely obedient, but might refuse to take actions that harm others or society, violate the law, etc.

These two axes -- level of autonomy and type of alignment -- are not entirely independent. For example, a sovereign passive system, while not quite self-contradictory, is a concept in tension, as is an obedient autonomous agent.\sidenote{We'll later see that while such systems buck the trend, that actually makes them very interesting and useful.} There's a clear sense in which autonomy and sovereignty tend to go hand-in-hand. In a similar vein, predictability tends to be higher in "passive" and "obedient" AI systems, whereas sovereign or autonomous ones will tend to be more unpredictable. All of this will be crucial for understanding the ramifications of potential AGI and superintelligence.

Creating truly aligned AI, of whatever flavor, requires solving three distinct challenges:

\begin{enumerate}
    \item Understanding what "we" want -- which is complex whether "we" means a specific person or organization (loyalty) or humanity broadly (sovereignty);
    \item Building systems that regularly act in accordance with those wants -- essentially creating consistent positive behavior;
    \item Most fundamentally, making systems that genuinely "care" about those wants rather than merely acting as if they do.
\end{enumerate}
The distinction between reliable behavior and genuine care is crucial. Just as a human employee might follow orders perfectly while lacking any real commitment to the organization's mission, an AI system might act aligned without truly valuing human preferences. We can train AI systems to say and do things through feedback, and they can learn to reason about what humans want. But making them {\em genuinely} value human preferences is a far deeper challenge.\sidenote[][-20mm]{This is not to say we require emotions or sentience. Rather, it is enormously difficult from the outside of a system to know what its inner goals, preferences, and values are. "Genuine" here would mean that we have strong enough reason to rely on it that in the case of critical systems we can bet our lives on it.}

The profound difficulties in solving these alignment challenges, and their implications for AI risk, will be explored further below. 
For now, understand that alignment is not just a technical feature we tack on to AI systems, but a fundamental aspect of their architecture that shapes their relationship with humanity.

\section{Key aspects of how modern general AI systems are made}

To really understand a human you need to know something about biology, evolution, child-rearing, and more; to understand AI you also need to know about how it is made. Over the past five years, AI systems have evolved tremendously both in capability and complexity. A key enabling factor has been the availability of very large amounts of computation (or colloquially "compute" when applied to AI). 

The numbers are staggering. About $10^{25}-10^{26}$ "floating-point operations" (FLOP)\sidenote[][-40mm]{$10^{25}$ means 1 followed by 25 zeros, or ten trillion trillion. A FLOP is just an arithmetic addition or multiplication of numbers with some precision. Note that AI hardware performance can vary by a factor of ten more depending upon the precision of the arithmetic and the architecture of the computer. Counting logic-gate operations (ANDS, ORS, 
AND NOTS) would be fundamental but these are not commonly available or 
benchmarked; for present purposes it is useful to standardize on 16-bit 
operations (FP16), though appropriate conversion factors should be 
established.} are used in the training of models like the GPT series, Claude, Gemini, etc.\sidenote[][8pt]{A collection of estimates and hard data is available from \href{https://epochai.org/data/large-scale-ai-models}{Epoch AI} and indicates about $2\times 10^{25}$ 16-bit FLOP for GPT-4; this roughly matches \href{https://mpost.io/gpt-4s-leaked-details-shed-light-on-its-massive-scale-and-impressive-architecture/}{numbers that were leaked} for GPT-4. Estimates for other mid-2024 models are all within a factor of a few of GPT-4.} 
(For comparison, if every human on Earth worked non-stop doing one calculation 
every five seconds, it would take around a billion years to accomplish this.)
This huge amount of computation enables training of models with up to trillions of model weights on terabytes of data -- a large fraction of all of the quality text that has ever been written alongside large libraries of sounds, images and video.
Complementing this training with additional extensive training reinforcing human preferences and good task performance, models trained in this way exhibit human-competitive performance across a significant span of basic intellectual tasks, including reasoning and problem solving. 

We also know (very, very roughly) how much computation speed, in operations per second, is sufficient for the {\em inference} speed\sidenote[][-15mm]{Inference is simply the process of generating an output from a neural network.  Training can be considered a succession of many inferences and model-weight tweaks.} of such a system to match the {\em speed} of human text processing. It is about $10^{15}-10^{16}$ FLOP per second.\sidenote[][-5mm]{For text production, the original GPT-4 required 560 TFLOP per token generated. Around 7 tokens/s is needed to keep up with human thought, so this gives $\approx 3\times 10^{15}\,$FLOP/s. But efficiencies have driven this down; \href{https://developer.nvidia.com/blog/supercharging-llama-3-1-across-nvidia-platforms/}{this NVIDIA brochure} for example indicates as little as $3\times 10^{14}$FLOP/s for a comparably-performing Llama 405B model.}
%

While powerful, these models are by their nature limited in key ways, quite analogous to how an individual human would be limited if forced to simply output text at a fixed rate of words per minute, without stopping to think or using any additional tools. 
More recent AI systems address these limitations through a more complex process and architecture 
combining several key elements:
\begin{itemize}
\item One or more neural networks, with one model providing the core cognitive capacity, and up to several others performing other more narrow tasks;
\item {\em Tooling} provided to and usable by the model -- for example ability to search the web, create or edit documents, execute programs, etc.
\item {\em Scaffolding} that connects input and outputs of neural networks.  A very simple scaffold might just allow two "instances" of an AI model to converse with each other, or one to check the work of another.\sidenote[][-45mm]{As a slightly more complex example, an AI system might first generate several possible solutions to a math problem, then use another instance to check each solution, and finally use a third to synthesize the results into a clear explanation. This allows for more thorough and reliable problem-solving than a single pass.}
\item {\em Chain-of-thought} and related prompting techniques do something similar, causing a model to for example generate many approaches to a problem, then process those approaches for an aggregate answer.
\item {\em Retraining} models to make better use of tools, scaffolding, and chain-of-thought.
\end{itemize}

Because these extensions can be very powerful (and include AI systems themselves), these composite systems can be quite sophisticated and dramatically enhance AI capabilities.\sidenote{See for example details on \href{https://openai.com/index/introducing-operator/}{OpenAI's "Operator"}, \href{https://docs.anthropic.com/en/docs/build-with-claude/computer-use}{Claude's tool capabilities}, and \href{https://github.com/Significant-Gravitas/AutoGPT}{AutoGPT}. OpenAI's \href{https://openai.com/index/introducing-deep-research/}{Deep Research} probably has a quite sophisticated architecture but details are not available.}
And recently, techniques in scaffolding and especially chain-of-thought prompting (and folding results back into retraining models to use these better) have been developed and employed in \href{https://openai.com/o1/}{o1}, \href{https://openai.com/index/openai-o3-mini/}{o3}, and \href{https://api-docs.deepseek.com/news/news250120}{DeepSeek R1} to do many passes of inference in response to a given query.\sidenote{Deepseek R1 relies on iteratively training and prompting the model so that the final trained model creates extensive chain-of-thought reasoning. Architectural details are not available for o1 or o3, however Deepseek has revealed that there is no particular "special sauce" required to unlock capability scaling with inference. But despite receiving a great deal of press as upending the "status quo" in AI, it does not impact the core claims of this essay.} This in effect allows the model to "think about" its response and dramatically boosts these models' ability to do high-caliber reasoning in science, math, and programming tasks.\sidenote[][7pt]{These models significantly outperform standard models on reasoning benchmarks. For instance, in the GPQA Diamond Benchmark—a rigorous test of PhD-level science questions—GPT-4o \href{https://openai.com/index/learning-to-reason-with-llms/}{scored} 56\%, while o1 and o3 achieved 78\% and 88\%, respectively, far exceeding the 70\% average score of human experts.}

For a given AI architecture, increases in training computation \href{https://arxiv.org/abs/2405.10938}{can be reliably translated} into improvements in a set of clearly-defined metrics. For less crisply defined general capabilities (such as those discussed below), the translation is less clear and predictive, but it is almost certain that larger models with more training computation will have new and better capabilities, even if it is hard to predict what those will be.

Similarly, composite systems and especially advances in "chain of thought" (and training of models that work well with it) have unlocked scaling in {\em inference} computation: for a given trained core model, at least some AI system capabilities increase as more computation is applied that allows them to "think harder and longer" about complex problems. This comes at a steep computing speed cost, requiring hundreds or thousands of more FLOP/s to match human performance.\sidenote{OpenAI's O3 probably expended $\sim 10^{21}-10^{22}\,$FLOP \href{https://www.interconnects.ai/p/openais-o3-the-2024-finale-of-ai}{to complete each of the ARC-AGI challenge questions}, which competent humans can do in (say) 10-100 seconds, giving a figure more like $\sim 10^{20}$\,FLOP/s.}

While only a part of what is leading to rapid AI progress,\sidenote[][1mm]{While computation is a key measure of AI system capability, it interacts with both data quality and algorithmic improvements. Better data or algorithms can reduce computational requirements, while more computation can sometimes compensate for weaker data or algorithms.} the role of computation and the possibility of composite systems will prove crucial to both preventing uncontrollable AGI and developing safer alternatives.

\section{What are AGI and superintelligence?}

The term "artificial general intelligence" has been around for some time to point to "human level" general-purpose AI. It has never been a particularly well-defined term, but in recent years it has paradoxically become no better defined yet even more important, with experts simultaneously arguing about whether AGI is decades away or already achieved, and trillion-dollar companies racing "to AGI." (The ambiguity of "AGI" was highlighted recently when \href{https://gizmodo.com/leaked-documents-show-openai-has-a-very-clear-definition-of-agi-2000543339}{leaked documents 
reportedly revealed} that in OpenAI's contract with Microsoft, AGI was defined as AI that 
generates \$100 billion in revenue for OpenAI -- a rather more mercenary than highbrow definition.)

There are two core problems with the idea of AI with "human level intelligence."  First, humans are very, very different in their ability to do any given type of cognitive work, so there is no "human level." Second, intelligence is very multi-dimensional; although there may be correlations, they are imperfect and may be quite different in AI. So even if we could define "human level" for many capabilities, AI would surely be far beyond it in some even while quite below in others.\sidenote{For instance, current AI systems far exceed human capability in rapid arithmetic or memory tasks, while falling short in abstract reasoning and creative problem-solving.}

It is, nonetheless, quite crucial to be able to discuss types, levels, and thresholds of AI capability. The approach taken here is to emphasize that general-purpose AI is here, and that it comes -- and will come -- at various capability levels at which it is convenient to attach terms even if they are reductive, because they correspond to crucial thresholds in terms of AI's effects on society and humanity.

We'll define "full" AGI to be synonymous with "super-human general-purpose AI" meaning an AI system that is able to perform essentially all human cognitive tasks at or above top human expert level, as well as acquire new skills and transfer capability to new domains. This is in keeping with how "AGI" is often defined in the modern literature. It's important to note that this is a {\em very} high threshold.  No human has this type of intelligence; rather it is the type of intelligence that large collections of top human experts would have if combined. We can term "superintelligence" a capability that goes beyond this, and define more limited levels of capability by "human-competitive" and "expert-competitive" GPAI, which perform a broad range of tasks at typical professional, or human expert level.\sidenote[][-50mm]{Very importantly, as a competitor such AI would have several major structural advantages including: it would not tire or have other individual needs like humans; it can be run at higher speeds just by scaling computing power; it can be copied along with any expertise or knowledge it acquires -- and neural networks' acquired knowledge can even be "merged" to transfer whole skillsets amongst themselves; it could communicate at machine speed; and it could self-modify or self-improve in more significant ways and higher speed than any human.}

These terms and some others are collected in \hyperref[tab:terms]{the table} below. 
For a more concrete sense of what the various grades of system can do, it is useful to take the definitions seriously and consider what they mean.

\renewcommand{\tablename}{}
\renewcommand{\thetable}{}
\definecolor{lightgreen}{RGB}{220,250,220}
\definecolor{lightyellow}{RGB}{255,255,224}
\definecolor{lightorange}{RGB}{255,223,186}
\definecolor{darkred}{RGB}{220,145,145}
\definecolor{lightred}{RGB}{255,204,203}
\definecolor{lightgray}{RGB}{231,231,231}

\makeatother
\begin{table}[ht]
    \footnotesize
    \renewcommand{\arraystretch}{1.2}
    \begin{tabular}{p{15mm} p{25mm} p{63mm} p{50mm}}
    \hline
    \textbf{AI Type} & \textbf{Related Terms} & \textbf{Definition} & \textbf{Examples} \\
    \hline
    Narrow AI & Weak AI & AI trained for a specific task or family of tasks. Excels in its domain but lacks general intelligence or transfer learning ability. & \parbox[t]{52mm}{Image recognition software;  Voice assistants (e.g., Siri, Alexa);  Chess-playing programs; DeepMind's AlphaFold} \\
    \hline
    Tool AI & Augmented Intelligence, AI Assistant & (Discussed later in essay.) AI system enhancing human capabilities. Combines human-competitive general-purpose AI, narrow AI, and guaranteed control, prioritizing safety and collaboration. Supports human decision-making. & \parbox[t]{52mm}{ Advanced coding assistants;  AI-powered research tools;  Sophisticated data analysis platforms. Competent but narrow and controllable agents} \\
    \hline
    General-purpose AI (GPAI) & & AI system adaptable to various tasks, including those not specifically trained for. & \parbox[t]{52mm}{ Language models (e.g., GPT-4, Claude);  Multimodal AI models;  DeepMind's MuZero} \\
    \hline
    Human-competitive GPAI & AGI [weak] & General-purpose AI performing tasks at average human level, sometimes exceeding it. & \parbox[t]{52mm}{ Advanced language models (e.g., O1, Claude 3.5);  Some multimodal AI systems} \\
    \hline
    Expert-competitive GPAI & AGI [partial] & General-purpose AI performing most tasks at human expert level, with significant but limited autonomy & Possibly a tooled and scaffolded O3, at least for mathematics, programming, and some hard sciences \\
    \hline
    AGI [full] & Super-human GPAI & AI system capable of autonomously performing roughly all human intellectual tasks at or beyond expert level, with efficient learning and knowledge transfer. & [No current examples - theoretical] \\
    \hline
    Super-intelligence & Highly super-human GPAI & AI system far surpassing human capabilities across all domains, outperforming collective human expertise. This out-performance could be in generality, quality, speed, and/or other measures. & [No current examples - theoretical] \\
    \hline
    \end{tabular}
    \begin{fullwidth}
        
        \mycaption{\flushleft Terms for different types and capability levels of AI systems used in this essay. These categories represent points along continuous spectra rather than sharp divisions.  Note that the term "AGI" is defined in particularly diverse ways.
        }
        \end{fullwidth}
    \label{tab:terms}
\end{table}

We're already experiencing what having GPAIs up to human competitive level is like. This has integrated relatively smoothly, as most users experience this as having a smart but limited temp worker who makes them more productive with mixed impact on the quality of their work.\sidenote{If you have not spent time using current top-of-the-line AI systems, I recommend it: they are genuinely useful and capable, and it is also important for calibrating the effect AI will have as they get more powerful.}

What would be different about expert-competitive GPAI is that it wouldn't have the core limitations of present-day AI, and would do the things experts do: independent economically valuable work, real knowledge creation, technical work you can count on, while rarely (though still occasionally) making dumb mistakes.

The idea of full AGI is that it {\em really does} all of the cognitive things even the most capable and effective humans do, autonomously and with no needed help or oversight. This includes sophisticated planning, learning new skills, managing complex projects, etc.  It could do original cutting-edge research.  It could run a company. Whatever your job is, if it is predominantly done by computer or over the phone, {\em it could do it at least as well as you.} And probably much faster and more cheaply.  We'll discuss some of the ramifications below, but for now the challenge for you is to really take this seriously. Imagine the top ten most knowledgeable and competent people you know or know of -- including CEOs, scientists, professors, top engineers, psychologists, political leaders, and writers. Wrap them all into one, who also speaks 100 languages, has a prodigious memory, operates quickly, is tireless and always motivated, and works at below minimum wage.\sidenote[][-60mm]{Consider a major research hospital: fully-realized AGI could simultaneously analyze all incoming patient data, keep up with every new medical paper, suggest diagnoses, design treatment plans, manage clinical trials, and coordinate staff scheduling -- all while operating at a level matching or exceeding the hospital's top specialists in each area. And it could do this for multiple hospitals simultaneously, at a fraction of the current cost. Unfortunately, you must also consider an organized crime syndicate: fully realized AGI could simultaneously hack, impersonate, spy on, and blackmail thousands of victims, keep up with law enforcement (which automates much more slowly), design new money-making schemes, and coordinate staff scheduling -- if there is any staff.}
That's a sense of what AGI would be.

For superintelligence the imagining is harder, because the idea is that it could perform intellectual feats that no human or even collection of humans can -- it is by definition unpredictable by us. But we can get a sense. As a bare baseline, consider lots of AGIs, each much more capable than even the top human expert, running at 100 times human speed, with enormous memory and terrific coordination capacity.\sidenote{In his \href{https://darioamodei.com/machines-of-loving-grace}{essay}, Dario Amodei, CEO of Anthropic, called to mind a "Country of [a million] geniuses".} And it goes up from there. Dealing with superintelligence would be less like conversing with a different mind, more like negotiating with a different (and more advanced) civilization.

So how close {\em are we} to AGI and superintelligence?

\section{At the threshold}

The past ten years have seen dramatic advances in AI driven by huge \href{https://epoch.ai/blog/training-compute-of-frontier-ai-models-grows-by-4-5x-per-year}{computational}, human, and \href{https://arxiv.org/abs/2405.21015}{fiscal} resources. Many narrow AI applications are better than humans at their assigned tasks, and are certainly far faster and cheaper.\sidenote[][-20mm]{You use a lot more of this AI than you probably think, driving speech generation and recognition, image processing, newsfeed algorithms, etc.} And there are also narrow super-human agents that can trounce all people at narrow-domain games such as \href{https://www.nature.com/articles/nature16961}{Go}, \href{https://arxiv.org/abs/1712.01815}{Chess}, and \href{https://www.deepstack.ai/}{Poker}, as well as more \href{https://deepmind.google/discover/blog/a-generalist-agent/}{general agents} that can plan and execute actions in simplified simulated environments as effectively as humans can.

Most prominently, current general AI systems from OpenAI/Micro\-soft, Google/Deepmind, Anthropic/Amazon, Facebook/Meta, X.ai/Tesla and others\sidenote[][-25mm]{While relationships between these pairs of companies are quite complex and nuanced, I have explicitly listed them to indicate both the vast overall market capitalization of firms now enjoined in AI development, and also that behind even "smaller" companies like Anthropic sit enormously deep pockets via investments and major partnership deals.} have emerged since early 2023 and steadily (though unevenly) increased their capabilities since then. All of these have been created via token-prediction on huge text and multimedia datasets, combined with extensive reinforcement feedback from humans and other AI systems.  Some of them also include extensive tool and scaffold systems.

\subsection*{Strengths and weaknesses of current general systems}

These systems perform well across an increasingly broad range of tests designed to measure intelligence and expertise, with progress that has surprised even experts in the field:
\begin{itemize}
    \item{When first released, GPT-4 \href{https://arxiv.org/abs/2303.08774}{matched or exceeded typical human performance} on standard academic tests including SATs, GRE, entrance exams, and bar exams. More recent models likely perform significantly 
    better, though results are not publicly available.}
    
    \item{The Turing test -- long considered a key benchmark for "true" AI -- is now routinely passed in some forms by modern language models, both informally and in \href{https://arxiv.org/abs/2405.08007}{formal studies}.\sidenote[][-180pt]{It has become fashionable to disparage the Turing test, but it is quite powerful and general. In weak versions it indicates whether typical people interacting with an AI (which is trained to act human) in typical ways for brief periods can tell whether it is an AI. They cannot. Second, a highly adversarial Turing test can probe essentially any element of human capability and intelligence -- by e.g. comparing an AI system to a human expert, evaluated by other human experts. There is a sense in which much of AI evaluation is a generalized form of Turing test.}}

    \item{On the comprehensive MMLU benchmark spanning 57 academic subjects, \href{https://paperswithcode.com/sota/multi-task-language-understanding-on-mmlu}{recent models 
    achieve domain-expert level scores} ($\sim 90\%$)\sidenote[][-13mm]{This is per domain - no 
    human could plausibly achieve such scores across all subjects simultaneously.}}
    
    \item{Technical expertise has advanced dramatically: The GPQA benchmark of 
    graduate-level physics saw \href{https://epoch.ai/data/ai-benchmarking-dashboard}{performance jump} from near-random guessing (GPT-4, 
    2022) to expert level (o1-preview, 2024).}
    
    \item{Even tests specifically designed to be AI-resistant are falling: 
    OpenAI's O3 \href{https://www.nextbigfuture.com/2024/12/openai-releases-o3-model-with-high-performance-and-high-cost.html}{reportedly} solves the ARC-AGI abstract problem-solving benchmark at 
    human level, achieves top-expert coding performance, and scores 25\% on 
    Epoch AI's "frontier math" problems designed to challenge elite 
    mathematicians.\sidenote[][-85pt]{These are problems that would take even excellent mathematicians substantial time to solve, if 
    they could solve them at all.}}
    
    \item{The trend is so clear that MMLU's developer has now created \href{https://agi.safe.ai/}{"Humanity's Last Exam"} -- an ominous name reflecting the possibility that AI will soon surpass human performance on any meaningful test. As of writing this, there are claims of AI systems achieving 27\% (according to \href{https://x.com/sama/status/1886220281565381078}{Sam Altman}) and 35\% (according to \href{https://arxiv.org/abs/2502.09955}{this paper}) on this extremely difficult exam. It is quite unlikely that any individual human could do so.
    }
\end{itemize}

Despite these impressive numbers (and their obvious intelligence when one interacts with them)\sidenote[][-180pt]{If you are of a skeptical bent, retain your skepticism but really take the most current models for a spin, as well as try for yourself some of the test questions they can pass. As a physics professor, I would predict with near certainty that, for example, the top models would pass the graduate qualifying exam in our department.} there are many things (at least the released versions of) these neural networks {\em cannot} do. 
Currently most are disembodied -- existing only on servers -- and process at most text, sound and still images (but not video.) Crucially, most cannot carry out complex planned activities requiring high accuracy.\sidenote[][-120pt]{This and other weaknesses like confabulation have slowed market adoption and led to a gap between perceived and claimed capabilities (which must also be viewed through the lens of intense market competition and the need to attract investment.) This has confused both the public and policymakers about the actual state of AI progress. While perhaps not matching the hype, the progress is very real.}
And there are a number of other qualities strong in high-level human cognition currently low in released AI systems.

The following \hyperref[tab:capabilities]{table} lists a number of these, based on mid-2024 AI systems such as GPT-4o, Claude 3.5 Sonnet, and Google Gemini 1.5.\sidenote[][-50pt]{The major advance since then has been development of systems trained for top-quality reasoning, leveraging more computation during inference and greater reinforcement learning. Because these models are new and their capabilities less tested, I've not wholly revamped this table except for "reasoning", which I regard as essentially solved. But I have updated predictions based on experienced and reported capabilities of those systems.} The key question for how rapidly general AI will become more powerful is: to what degree will just doing {\em more of the same} produce results, versus adding additional but {\em known} techniques, versus developing or implementing {\em really new} AI research directions. My own predictions for this are given in the table, in terms of how likely each of these scenarios is to get that capability to and beyond human level.
\clearpage

\begin{table}[ht]
    \footnotesize
    \begin{adjustwidth}{-0.8cm}{0cm}
    \renewcommand{\arraystretch}{1.5}
    \begin{tabular}{p{16mm} p{63mm} p{69mm} p{8mm}}
    \hline
    Capability & Description of capability & Status/prognosis & 
    \multicolumn{1}{c}{Scaling/} \\
    & & & \multicolumn{1}{c}{known/new} \\
    \hline
    \multicolumn{4}{l}{\textit{Core Cognitive Capabilities}} \\
    Reasoning & People can do accurate, multistep reasoning, following rules and 
    checking accuracy. & Dramatic recent progress using extended chain-of-thought and retraining & 95/5/5 \\[1ex]
    
    Planning & People exhibit long-term and hierarchical planning. & Improving with 
    scale; can be strongly aided using scaffolding and better training techniques. & 
    10/85/5 \\[1ex]
    
    Truth-grounding & GPAIs confabulate ungrounded information to satisfy queries. & 
    Improving with scale; calibration data available within model; can be 
    checked/improved via scaffolding. & 30/65/5 \\[1ex]
    
    Flexible problem-solving & Humans can recognize new patterns and invent new 
    solutions to complex problems; current ML models struggle. & Improves with scale 
    but weakly; may be solvable with neurosymbolic or generalized "search" 
    techniques. & 15/75/10 \\[3ex]
    \hline
    \multicolumn{4}{l}{\textit{Learning and Knowledge}} \\
    Learning \& memory & People have working, short-term, and long-term memory, all 
    of which are dynamic and inter-related. & All models learn during training; GPAIs 
    learn within context window and during fine-tuning; "continual learning" and 
    other techniques exist but not yet integrated into large GPAIs. & 5/80/15 \\[1ex]
    
    Abstraction \& recursion & People can map and transfer relation sets into more abstract ones 
    for reasoning and manipulation, including recursive "meta" reasoning. & Weakly 
    improving with scale; could emerge in neurosymbolic systems. & 30/50/20 \\[1ex]
    
    World model(s) & People have and continually update a predictive world model within which they can solve problems and do physical reasoning & 
    Improving with scale; updating tied to learning; GPAIS weak in real-world 
    prediction. & 20/50/30 \\[3ex]

    \hline
    \multicolumn{4}{l}{\textit{Self and Agency}} \\
    Agency & People can take actions in order to pursue goals, based on 
    planning/prediction. & Many ML systems are agentic; LLMs can be made agents via 
    wrappers. & 5/90/5 \\[1ex]
    
    Self-direction & People develop and pursue their own goals, with 
    internally-generated motivation and drive. & Largely composed of agency plus 
    originality; likely to emerge in complex agential systems with abstract goals. & 
    40/45/15 \\[1ex]
    
    Self-reference & People understand and reason about themselves as situated within 
    an environment/context. & Improving with scale and could be augmented with 
    training reward. & 70/15/15 \\[1ex]
    
    Self-awareness & People have knowledge of and can reason regarding their own 
    thoughts and mental states. & Exists in some sense in GPAIs, which can arguably 
    pass the classic "mirror test" for self-awareness. Can be improved with 
    scaffolding; but unclear if this is enough. & 20/55/25 \\[3ex]
    \hline
    \multicolumn{4}{l}{\textit{Interface and Environment}} \\
    Embodied intelligence & People understand and actively interact with their 
    real-world environment. & Reinforcement learning works well in simulated and 
    real-world (robotic) environments and can be integrated into multimodal 
    transformers. & 5/85/10 \\[1ex]
    
    Multi-sense processing & People integrate and real-time process visual, audio, 
    and other sensory streams. & Training in multiple modalities appears to "just 
    work," and improve with scale. Realtime video processing is difficult but 
    e.g. self-driving systems are rapidly improving. & 30/60/10 \\[3ex]

    \hline
    \multicolumn{4}{l}{\textit{Higher-order Capabilities}} \\
    Originality & Current ML models are creative in transforming and combining 
    existing ideas/works, but people can build new frameworks and structures, 
    sometimes tied to their identity. & Can be hard to discern from "creativity," 
    which may scale into it; may emerge from creativity plus self-awareness. & 50/40/10 \\[1ex]
    
    Sentience & People experience qualia; these can be positive, negative or neutral 
    valence; it is "like something" to be a person. & Very difficult and 
    philosophically fraught to determine whether a given system has this. & 5/10/85 \\
    \hline
    \end{tabular}

       \begin{fullwidth}
\mycaption{\flushleft Key capabilities currently below human expert level in 
modern GPAI systems, grouped by type. The third column summarizes current status. 
Final column shows predicted likelihood (\%) that human-level performance will be 
achieved through: scaling current techniques / combining with known techniques / 
developing new techniques. These capabilities are not independent, and increase in 
any one typically goes along with increases in others. Note that not all (particularly sentience) are 
necessary for AI systems capable of advancing AI development, highlighting the 
possibility of powerful but non-sentient AI.}
\end{fullwidth}
\end{adjustwidth}
   \label{tab:capabilities}
\end{table}

\clearpage

Breaking down what is "missing" in this way makes it fairly clear that we are quite on-track for broadly above-human intelligence by scaling existing or known techniques.\sidenote[][]{Previous waves of AI optimism in the 1960s and 1980s ended in "AI winters" when promised capabilities failed to materialize. However, the current wave differs fundamentally in having achieved superhuman performance in many domains, backed by massive computational resources and commercial success.}  

There could still be surprises.  Even putting aside "sentience," there could be some of the listed core cognitive capabilities that really can't be done with current techniques and require new ones.  But consider this. The present effort being put forth by many of the world's largest companies amounts to multiple times the Apollo project's and tens of times the Manhattan project's spend,\sidenote[][5mm]{The full Apollo project \href{https://www.planetary.org/space-policy/cost-of-apollo}{cost about \$250bn USD in 2020 dollars}, and the Manhattan project \href{https://www.brookings.edu/the-costs-of-the-manhattan-project/}{less than a tenth that}. Goldman Sachs \href{https://www.datacenterdynamics.com/en/news/goldman-sachs-1tn-to-be-spent-on-ai-data-centers-chips-and-utility-upgrades-with-little-to-show-for-it-so-far/}{projects a trillion dollars of spend just on AI data centers} over the next few years.} and is employing thousands of the very top technical people at unheard of salaries.  The dynamics of the past few years have now brought to bear more human intellectual firepower (with AI now being added) to this than any endeavor in history.  We should not bet on failure.

\subsection*{The big target: generalist autonomous agents}

The development of general AI over the past several years has focused on creating 
general and powerful but tool-like AI: it functions primarily as a (fairly) loyal 
assistant, and generally does not take actions on its own. This is partly by 
design, but largely because these systems have simply not been competent enough at 
the relevant skills to be entrusted with complex actions.\sidenote[][-30mm]{Although humans make plenty of mistakes, we underestimate just how reliable we can be!  Because probabilities multiply, a task requiring 20 steps to do correctly requires each step to be 97\% reliable just to get it done right half the time. We do such tasks all the time.}

AI companies and researchers are, however, increasing \href{https://www.axios.com/2025/01/23/davos-2025-ai-agents}{shifting focus} 
toward {\em autonomous} expert-level general-purpose agents.\sidenote{A strong move in this direction has very recently been taken with OpenAI's \href{https://openai.com/index/introducing-deep-research/}{"Deep Research"} assistant that autonomously performs general research, described as "a new agentic capability that conducts multi-step research on the internet for complex tasks."} This would allow the systems to act more like a human assistant to which the user can delegate real actions.\sidenote[][10mm]{Things like fill in that pesky PDF form, book flights, etc. But with a PhD in 20 fields! So also: write that thesis for you, negotiate that contract for you, prove that theorem for you, create that ad campaign for you, etc. What do {\em you} do? You tell it what to do, of course.}
What will that take?  A number of the capabilities in the "what's missing" table are implicated, including strong truth-grounding, learning and memory, abstraction and recursion, and world-modeling (for intelligence), planning, agency, originality, self-direction, self-reference, and self-awareness (for autonomy), and multi-sense-processing, embodied intelligence, and flexible problem-solving (for generality).\sidenote[][]{Note that sentience is {\em not} clearly required, nor does AI in this triple-intersection necessarily imply it.} 

This triple-intersection of high autonomy (independence of action), high generality (scope and task breadth) and high intelligence (competence at cognitive tasks) is currently unique to humans.
It is implicitly what many probably have in mind when they think 
of AGI -- both in terms of its value as well as its risks. 

This provides another way to define A-G-I as {\bf\em A}utonomous-{\bf\em G}eneral-{\bf\em I}ntelligence, and we'll see that this triple intersection provides a very valuable lens for high-capability systems both in understanding their risks and rewards, and in governance of AI.

\begin{figure*}[!ht]
\centering
\includegraphics[width=0.85\linewidth]{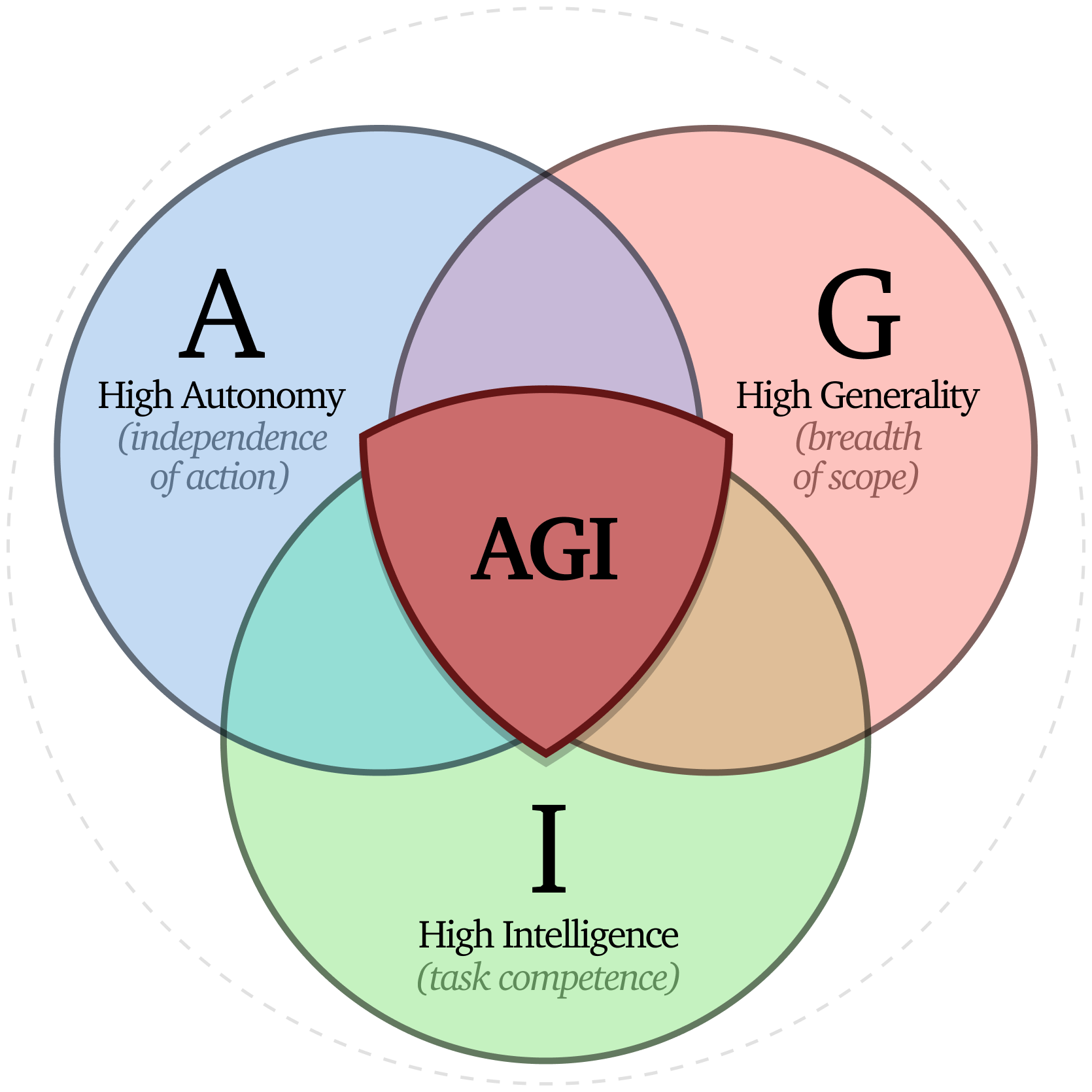}
    \begin{fullwidth}
    
\mycaption{\small The transformative A-G-I power and risk zone emerges from the intersection of three key properties: high Autonomy,\\ high Intelligence at tasks, and high Generality.}
\end{fullwidth}
\label{fig:agi-venn}
\end{figure*}

\subsection*{The AI (self-)improvement cycle}
\label{sec:self-improve}

A final crucial factor in understanding AI progress is AI's unique technological feedback loop.  In developing AI, success -- in both demonstrated systems and deployed products -- brings additional investment, talent, and competition, and we are currently in the midst of an enormous AI hype-plus-reality feedback loop that is driving hundreds of billions, or even trillions, of dollars in investment.
 
This type of feedback cycle could happen with any technology, and we've seen it in many, where market success begets investment, which begets improvement and better market success. But AI development goes further, in that now AI systems are helping to develop new and more powerful AI systems.\sidenote[][-23pt]{A close analogy here is chip technology, where development has maintained Moore's law for decades, as computer technologies help people design the next generation of chip technology. AI will be far more direct.}
We can think of this feedback loop in five stages, each with a shorter timescale than the last, as shown in the \hyperref[tab:improvement-cycle]{table}.

\begin{table}[ht]
    \small
    \renewcommand{\arraystretch}{1.2}
    \begin{tabular}{p{20mm} p{15mm} p{45mm} p{40mm} p{25mm}}
    \hline
    \textbf{Stage} & \textbf{Timescale} & \textbf{Key Drivers} & 
    \textbf{Current Status} & \textbf{Rate-Limiting Factors} \\
    \hline
    Infrastructure & Years & AI success → investment → better hardware/infrastructure 
    & Ongoing; massive\newline investment & Hardware\newline development cycle \\
    \hline
    Model\newline Development & 1-2 Years & Human-led research with AI assistance & 
    Active across major labs & Training run complexity \\
    \hline
    Data \newline Generation & Months & AI systems generating\newline synthetic training data & 
    Beginning phase & Data quality verification \\
    \hline
    Tool \newline Development & Days Weeks & AI systems creating their own scaffolding/tools & 
    Early experiments & Software\newline integration time \\
    \hline
    Network Self Improvement & Hours weeks & Groups of AI systems innovate "social" institutions & 
    Unknown & Inference rate \\
    \hline
    Recursive Improvement & Unknown & AGI/superintelligent systems autonomously self-improving & 
    Not yet possible & Unknown/\newline unpredictable\\
    \hline
    \end{tabular}
          \begin{fullwidth}
    \mycaption{\small\\ The AI improvement cycle operates across multiple timescales, with each stage potentially accelerating subsequent stages. Earlier stages are well underway, while later stages remain speculative but could proceed very rapidly once unlocked.}
    \end{fullwidth}
    \label{tab:improvement-cycle}
\end{table}

Several of these stages are already underway, and a couple clearly getting started. The last stage, in which AI systems autonomously improve themselves, has been a staple of the literature on the risk of very powerful AI systems, and for good reason.\sidenote{It's important to let it sink in for a moment that AI could -- soon -- be improving itself on a timescale of days or weeks. Or less. Keep this in mind when someone tells you an AI capability is definitely far away.} But it is important to note that it is just the most drastic form of a feedback cycle that has already started and could lead to more surprises in the rapid advancement of the technology.

\section{The race for AGI}

The recent fast progress in AI has resulted both from and in an extraordinary level of attention and investment.  This is driven in part by success in AI development, but more is going on. Why are some of the largest companies on Earth, and even countries, racing to build not just AI, but AGI and superintelligence?

\subsection*{What has driven AI research toward human-level AI}

Until the past five years or so, AI has been largely an academic and scientific research problem, thus largely driven by curiosity and the drive to understand intelligence and how to create it in a new substrate.  

In this phase, there was relatively little attention paid to the benefits or perils of AI among most researchers.  When asked why AI should be developed, a common response might be to list, somewhat vaguely, problems that AI could help with: new medicines, new materials, new science, smarter processes, and in general improving things for people.\sidenote{A more precise list of worthy goals is the UN \href{https://sdgs.un.org/goals}{Sustainable Development Goals.} These are, in a sense, the closest we have to a set of global consensus goals for what we'd like to see improved in the world. AI could help.}

These are admirable goals!\sidenote[][20mm]{Technology in general has a transformative economic and social power for human betterment, as thousands of years attest. In this vein, a long and compelling explication of a positive AGI vision can be found in \href{https://darioamodei.com/machines-of-loving-grace}{this essay} by Anthropic founder Dario Amodei.} Although we can and will question whether AGI -- rather than AI in general -- is necessary for these goals, they exhibit the idealism with which many AI researchers started.

Over the past half-decade, however, AI has transformed from a relatively pure research field into much more of an engineering and product field, largely driven by some of the world's largest companies.\sidenote[][10mm]{Private AI investment \href{https://cset.georgetown.edu/publication/tracking-ai-investment/}{started to boom in 2018-19, crossing public investment around then,} and has hugely outpaced it since.}  Researchers, while relevant, are no longer in charge of the process.

\subsection*{Why are companies trying to build AGI?}

So why are giant corporations (and even more so investors) pouring vast resources into building AGI? There are two drivers that most companies are quite honest about: they see AI as drivers of productivity for society, and of profits for them. Because general AI is by nature general-purpose, there is a huge prize: rather than choosing a sector in which to create products and services, one can try {\em all of them at once.} Big Tech companies have grown enormous by producing digital goods and services, and at least some executives surely see AI as simply the next step in providing them well, with risks and benefits that expand upon but echo those provided by search, social media, laptops, phones, etc.

But why AGI? There is a very simple answer to this, which most companies and investors shy away from discussing publicly.\sidenote[][-20mm]{I can attest that behind more closed doors, they have no such compunction. And it's becoming more public; see for example Y-combinator's new \href{https://www.ycombinator.com/rfs}{"request for startups"}, many parts of which explicitly call for wholesale replacement of human workers. To quote them, "The value prop of B2B SaaS was to make human workers incrementally more efficient.  The value prop of vertical AI agents is to automate the work entirely...It's entirely possible this opportunity is big enough to mint another 100 unicorns." (For those not versed in Silicon Valley speak, "B2B" is business-to-business and a unicorn is a \$1 billion company. That is they are talking about more than a hundred billion-plus-dollar businesses that replace workers for other businesses.)}

It is that AGI can directly, one-for-one, {\em replace workers.}

Not augment, not empower, not make more productive. Not even {\em displace.} All of these can and will be done by non-AGI.  AGI is specifically what can fully {\em replace} thought workers (and with robotics, many physical ones as well.)
As support for this view one need look no further than OpenAI's \href{https://openai.com/our-structure/}{(publicly stated) definition} of AGI, which is "a highly autonomous system that outperforms humans at most economically valuable work." 

The prize here (for companies!) is enormous. Labor costs are a substantial percentage of the world's $\sim \$100\,$trillion global economy. Even if only a fraction of this is captured by replacement of human labor by AI labor, this is trillions of dollars of annual revenue. AI companies are also cognizant of who is willing to pay.  As they see it, you are not going to pay thousands of dollars a year for productivity tools.  But a company {\em will} pay thousands of dollars per year to replace your labor, if they can.

\subsection*{Why countries feel they have to race to AGI}

Countries' stated motivations for pursuing AGI focus on economic and scientific 
leadership. The argument is compelling: AGI could dramatically accelerate 
scientific research, technological development, and economic growth. Given the 
stakes, they argue, no major power can afford to fall behind.\sidenote{See for example a recent \href{https://www.uscc.gov/sites/default/files/2024-11/2024_Executive_Summary.pdf}{US-China Economic and Security Review Commission report}. Although there was surprisingly little justification within the report itself, the top-line recommendation  was that the US "Congress establish and fund a Manhattan Project-like program dedicated to racing to and acquiring an Artificial General Intelligence (AGI) capability."}

But there are also additional and largely unstated drivers. There is no doubt that when certain military and national security leaders meet behind closed doors to discuss an extraordinarily 
potent and catastrophically risky technology, their focus is not on "how do we 
avoid those risks" but rather "how do we get this first?"  Military and intelligence leaders see AGI as a potential revolution in military affairs, perhaps the most significant since nuclear weapons. The fear is that the first country to develop AGI could gain an insurmountable strategic advantage. This creates a classic arms race dynamic.

We'll see that this "race to AGI" thinking,\sidenote{Companies are now adopting this geopolitical framing as a shield against any constraint on their AI development, generally in ways that are blatantly self-serving, and sometimes in ways that don't even make basic sense. Consider Meta's \href{https://about.fb.com/news/2025/02/meta-approach-frontier-ai/}{Approach to Frontier AI}, which simultaneously argues that America must "[Cement its] position as a leader in technological innovation, economic growth and national security" and also that it must do so by openly releasing its most powerful AI systems -- which includes giving them directly to its geopolitical rivals and adversaries.}
 while compelling, is deeply flawed. This is not because racing is dangerous and risky -- though it is -- but due to the nature of the technology. The unstated assumption is that AGI, like other technologies, is controllable by the state that develops it, and is a power-granting boon to the society that has the most of it. As we will see, it probably won't be either.

\subsection*{Why superintelligence?}

While companies publicly focus on productivity, and countries on economic and technological growth, for those deliberately pursuing full AGI and superintelligence these are just the start. What do they really have in mind?  Although seldom said out loud, they include:

\begin{enumerate}
\item Cures for many or all diseases;
\item Stopping and reversal of aging;
\item New sustainable energy sources like fusion;
\item Human upgrades, or designer organisms via genetic engineering;
\item Nanotechnology and molecular manufacturing;
\item Mind uploads;
\item Exotic physics or space technologies;
\item Super-human advice and decision-support;
\item Super-human planning and coordination.
\end{enumerate}

The first three are largely "single-edge" technologies – i.e. likely to be quite strongly net positive. It's hard to argue against curing diseases or being able to live longer if one chooses. And we have already reaped the negative side of fusion (in the form of nuclear weapons); it would be lovely now to get the positive side. The question with this first category is whether getting these technologies sooner compensates for the risk.

The next four are clearly double-edged: transformative technologies with both potentially huge upsides and immense risks, much like AI. All of these, if they sprung out of a black-box tomorrow and were deployed, would be incredibly difficult to manage.\sidenote{Thus we'd likely have to leave management of these technologies to the AIs. But this would be a very problematic delegation of control, which we'll return to below.}

The final two concern the super-human AI doing things itself rather than just inventing technology. More precisely, putting euphemisms aside, these involve powerful AI systems telling people what to do. Calling this "advice" is disingenuous if the system doing the advising is far more powerful than the advised, who cannot meaningfully understand the basis of decision (or even if this is provided, trust that the advisor would not provide a similarly compelling rationale for a different decision.)

This points to a key item missing from the above list:
\begin{enumerate}
\setcounter{enumi}{9}
   \item Power.
\end{enumerate}
It is abundantly clear that much of what is underlying the current race for super-human AI is the idea that {\em intelligence = power}. Each racer is banking on being the best holder of that power, and that they will be able to wield it for ostensibly benevolent reasons without it slipping or being taken from their control.

That is, what companies and nations are really chasing is not just the fruits 
of AGI and superintelligence, but the power to control who gets access to them and how they're used. Companies see themselves as responsible stewards of this power in service of shareholders and humanity; nations see themselves as necessary 
guardians preventing hostile powers from gaining decisive advantage. Both are dangerously wrong, failing to recognize that superintelligence, by its 
nature, cannot be reliably controlled by any human institution. We will see that the nature and dynamics of superintelligent systems make human control extremely difficult, if not impossible.

These racing dynamics -- both corporate and geopolitical -- make certain risks nearly inevitable unless decisively interrupted. We turn now to examining these risks and why they cannot be adequately mitigated within a competitive\sidenote[][]{Competition in technology development often brings important benefits: preventing monopolistic control, driving innovation and cost reduction, enabling diverse approaches, and creating mutual oversight. However, with AGI these benefits must be weighed against unique risks from racing dynamics and pressure to reduce safety precautions.} development paradigm.

\section{What happens if we build AGI on our current path?}

The development of full artificial general intelligence -- what we will call here AI that is "outside the Gates" -- would be a fundamental shift in the nature of the world: by its very nature it means adding a new species of intelligence to Earth with greater capability than that of humans. 

What will then happen depends on many things, including the nature of the technology, choices by those developing it, and the world context in which it is being developed.

Currently, full AGI is being developed by a handful of massive private companies in a race with each other, with little meaningful regulation or external oversight,\sidenote{The \href{https://artificialintelligenceact.eu/}{EU AI act} is a significant piece of legislation but would not directly prevent a dangerous AI system from being developed or deployed, or even openly released, especially in the US. Another significant piece of policy, the US Executive order on AI, has been rescinded.} in a society with increasingly weak and even dysfunctional core institutions,\sidenote[][20pt]{This \href{https://news.gallup.com/poll/1597/confidence-institutions.aspx}{Gallup poll} shows a bleak decline in trust in public institutions since 2000 in the US. European numbers are varied and less extreme, but also on a downward trend. Distrust does not strictly mean institutions really {\em are} dysfunctional, but it is an indication as well as a cause.} in a time of high geopolitical tension and low international coordination. Although some are altruistically motivated, many of those doing it are driven by money, or power, or both.

Prediction is very difficult, but there are some dynamics that are well enough understood, and apt-enough analogies with previous technologies to offer a guide. And unfortunately, despite AI's promise, they give good reason to be profoundly pessimistic about how our current trajectory will play out.

To put it bluntly, on our present course developing AGI will have some positive effects (and make some people very, very rich). But the nature of the technology, the fundamental dynamics, and the context in which it is being developed, strongly indicate that: powerful AI will dramatically undermine our society and civilization; we will lose control of it; we may well end up in a world war because of it; we will lose (or cede) control {\em to} it; it will lead to artificial superintelligence, which we absolutely will not control and will mean the end of a human-run world.

These are strong claims, and I wish they were idle speculation or unwarranted "doomer"ism. But this is where the science, the game theory, the evolutionary theory, and history all point. This section develops these claims, and their support, in detail.

\subsection*{We will undermine our society and civilization}

Despite what you may hear in Silicon Valley boardrooms, most disruption -- especially of the very rapid variety -- is not beneficial. There are vastly more ways to make complex systems worse than better. Our world functions as well as it does because we have painstakingly built processes, technologies, and institutions that have made it steadily better.\sidenote{And major disruptions we now endorse -- such as expansion of rights to new groups -- were specifically driven by people in a direction towards making things better.}
Taking a sledgehammer to a factory rarely improves operations.

Here is an (incomplete) catalog of ways AGI systems would disrupt our civilization. 

\begin{itemize}
\item {They would dramatically disrupt labor, leading {\em at bare minimum} to 
dramatically higher income inequality and potentially large-scale under-employment 
or unemployment, on a timescale far too short for society to 
adjust.\sidenote[][-10mm]{Let me be blunt. If your job can be done from behind a 
computer, with relatively little in-person interaction with people outside of 
your organization, and does not entail legal responsibility to external parties, 
it would by definition be possible (and likely cost-saving) to completely swap you out 
for a digital system. Robotics to replace much physical labor will come later -- but not that much later once AGI starts designing robots.}}

\item {They would likely lead to the concentration of vast economic, social, and 
political power -- potentially more than that of nation states -- into a small 
number of massive private interests unaccountable to the public.}

\item {They could suddenly make previously difficult or expensive activities 
trivially easy, destabilizing social systems that depend on certain activities 
remaining costly or requiring significant human effort.\sidenote{For example, what happens to our judicial system if lawsuits are nearly-free to file? What happens when bypassing security systems through social engineering becomes cheap, easy, and risk-free?}}

\item {They could flood society's information gathering, processing, and 
communication systems with completely realistic yet false, spammy, 
overly-targeted, or manipulative media so thoroughly that it becomes 
impossible to tell what is physically real or not, human or not, factual or not, 
and trustworthy or not.\sidenote{\href{https://www.linkedin.com/pulse/projected-growth-ai-generated-data-public-internet-our-arun-kumar-r-vhije/}{This article } claims that 10\% of all internet content is already AI-generated, and is Google's top hit (for me) to the search query "estimates of what fraction of new internet content is AI-generated."  Is it true?  I have no idea! It cites no references and it wasn't written by a person.  What fraction of new images indexed by Google, or Tweets, or comments on Reddit, or Youtube videos are generated by humans? Nobody knows -- I don't think it is a knowable number. And this less than {\em two years} into the advent of generative AI.}}

\item {They could create dangerous and near total intellectual dependence, where human 
understanding of key systems and technologies atrophies as we increasingly rely 
on AI systems we cannot fully comprehend.}

\item {They could effectively end human culture, once nearly all cultural objects 
(text, music, visual art, film, etc.) consumed by most people are created, 
mediated, or curated by nonhuman minds.}

\item {They could enable effective mass surveillance and manipulation systems 
usable by governments or private interests to control a populace and pursue 
objectives in conflict with the public interest.}

\item {By undermining human discourse, debate, and election systems, they could 
reduce the credibility of democratic institutions to the point where they are 
effectively (or explicitly) replaced by others, ending democracy in states where 
it currently exists.}

\item {They could become, or create, advanced self-replicating intelligent 
software viruses and worms that could proliferate and evolve, massively 
disrupting global information systems.}

\item {They can dramatically increase the ability of terrorists, bad actors, and 
rogue states to cause harm via biological, chemical, cyber, autonomous, or other 
weapons, without AI providing a counterbalancing ability to prevent such harm. Similarly they would undermine national security and geopolitical balances by making top-tier nuclear, bio, engineering, and other expertise available to regimes that would not otherwise have it.}

\item {They could cause rapid large-scale runaway hyper-capitalism, with 
effectively AI-run companies competing in largely electronic financial, sales, 
and services spaces. AI-driven financial markets could operate at speeds and 
complexities far beyond human comprehension or control. All of the failure modes 
and negative externalities of current capitalist economies could be exacerbated 
and sped far beyond human control, governance, or regulatory capability.}

\item {They could fuel an arms race between nations in AI-powered weaponry, 
command-and-control systems, cyberweapons, etc., creating very rapid buildup of 
extremely destructive capabilities.}
\end{itemize}

These risks are not speculative. Many of them are being realized as we speak, via existing AI systems! But consider, {\em really} consider, what each would look like with dramatically more powerful AI. 

Consider labor displacement when most workers simply cannot provide any significant economic value beyond what AI can, in their field of expertise or experience -- or even if they retrain!
Consider mass surveillance if everyone is being individually watched and monitored by something faster and cleverer than themselves. What does democracy look like when we cannot reliably trust any digital information that we see, hear, or read, and when the most convincing public voices are not even human, and have no stake in the outcome? What becomes of warfare when generals have to constantly defer to AI (or simply put it in charge), lest they grant a decisive advantage to the enemy? Any one of the above risks represents a catastrophe for human\sidenote[][-10mm]{Also worth adding is that there is "moral" risk that we might create digital beings that can suffer.  As we currently do not have a reliable theory of consciousness that would allow us to distinguish physical systems that can and cannot suffer, we cannot rule this out theoretically.  Moreover, AI systems' reports of their sentience are likely unreliable with respect to their actual experience (or non-experience) of sentience.} civilization if fully realized.

You can make your own predictions. Ask yourself these three questions for each risk:
\begin{enumerate}
\item Would super-capable, highly autonomous, and very general AI allow it in a way or at a scale that would not otherwise be possible?
\item Are there parties who would benefit from things that cause it to happen?
\item Are there systems and institutions in place that would effectively prevent it from happening?
\end{enumerate}

Where your answers are "yes, yes, no" you can see we have got a big problem.

What is our plan for managing them? As it stands there are two on the table regarding AI in general.

The first is to build safeguards into the systems to prevent them from doing things they shouldn't. That's being done now: commercial AI systems will, for example, refuse to help build a bomb or write hate speech.

This plan is woefully inadequate for systems outside the Gate.\sidenote{Technical solutions in this field of AI "alignment" are unlikely to be up to the task either. In present systems they work at some level, but are shallow and can generally be circumvented without significant effort; and as discussed below we have no real idea how to do this for much more advanced systems.} It may help decrease risk of AI providing manifestly dangerous assistance to bad actors. But it will do nothing to prevent labor disruption, concentration of power, runaway hyper-capitalism, or replacement of human culture: these are just results of using the systems in permitted ways that profit their providers! And governments will surely obtain access to systems for military or surveillance use. 

The second plan is even worse: simply to openly release very powerful AI systems for anyone to use as they like,\sidenote{Such AI systems may come with some built-in safeguards. But for any model with anything like current architecture, if full access to its weights are available, safety measures can be stripped away via additional training or other techniques. So it is virtually guaranteed that for each system with guardrails there will also be a widely available system without them.  Indeed Meta's Llama 3.1 405B model was openly released with safeguards.  But {\em even before that} a "base" model, with no safeguards, was leaked.}
and hope for the best. 

Implicit in both plans is that someone else, e.g. governments, will help to solve the problems through soft or hard law, standards, regulations, norms, and other mechanisms we generally use to manage technologies.\sidenote[][15mm]{Could the market manage these risks without government involvement?  In short, no. There are certainly risks that companies are strongly incentivized to mitigate. But many others companies can and do externalize to everyone else, and many of the above are in this class: there are no natural market incentives to prevent mass surveillance, truth decay, concentration of power, labor disruption, damaging political discourse, etc.  Indeed we have seen all of these from present-day tech, especially social media, which has gone essentially unregulated. AI would just hugely amp up many of the same dynamics.} But putting aside that AI corporations already fight tooth-and-nail against any substantial regulation or externally imposed limitations at all, for a number of these risks it's quite hard to see what regulation would even really help. Regulation could impose safety standards on AI. But would it prevent companies from replacing workers wholesale with AI? Would it forbid people from letting AI run their companies for them? Would it prevent governments from using potent AI in surveillance and weaponry? These issues are fundamental. Humanity could potentially find ways to adapt to them, but only with {\em much} more time. As it is, given the speed that AI is reaching or exceeding the capabilities of the people trying to manage them, these problems look increasingly intractable.

\vspace{-0.25cm}
\subsection*{We will lose control of (at least some) AGI systems}
\label{sec:losecontrolof}
\vspace{-0.1cm}

Most technologies are very controllable, by construction. If your car or your toaster starts doing something you don't want it to do, that's just a malfunction, not part of its nature as a toaster. AI is different: it is {\em grown} rather than designed, its core operation is opaque, and it is inherently unpredictable. 

This loss of control isn't theoretical -- we see early versions already. Consider first a prosaic, and arguably benign example. If you ask ChatGPT to help you mix a poison, or write a racist screed, it will refuse. That's arguably good. But it is also ChatGPT {\em not doing what you've explicitly asked it to do}. Other pieces of software do not do that. That same model won't design poisons at the request of an OpenAI employee either.\sidenote{OpenAI likely has more obedient models for internal use. It's unlikely that OpenAI has built some sort of "backdoor" so that ChatGPT can be better controlled by OpenAI itself, because this 
would be a terrible security practice, and be highly exploitable given AI's 
opacity and unpredictability.} This makes it very easy to imagine what it would be like for future more powerful AI to be out of control.  In many cases, they will simply not do what we ask! Either a given super-human AGI system will be absolutely obedient and 
loyal to some human command system, or it won't. If not, {\em it will do things 
it may believe are good for us, but that are contrary to our explicit commands.} 
That isn't something that is under control. 
But, you might say, this is intentional -- these refusals are by design, part of what is called "aligning" the systems to human values. And this is true. However the alignment "program" itself has two major problems.\sidenote[][-20pt]{Also of crucial importance: alignment or any other safety features only matter if they are actually used in an AI system. Systems that are openly released (i.e. where model weights and architecture are publicly available) can be transformed relatively easily into systems {\em without} those safety measures. Open-releasing smarter-than-human AGI systems would be astonishingly reckless, and it is hard to imagine how human control or even relevance would be maintained in such a scenario. There would be every motivation, for example, to let loose powerful self-reproducing and self-sustaining AI agents with the goal to make money and send it to some cryptocurrency wallet. Or to win an election. Or overthrow a government. Could "good" AI help contain this? Perhaps -- but only by delegating huge authority to it, leading to control loss as described below.}

First, at a deep level we have no idea how to do it. How do we guarantee that an AI system will "care" about what we want? We can train AI systems to say and not say things by providing feedback; and they can learn and reason about what humans want and care about just as they reason about other things. But we have no method -- even theoretically -- to cause them to deeply and reliably value what people care about. There are high-functioning human psychopaths who know what is considered right and wrong, and how they are supposed to behave.  They simply don't {\em care}. But they can {\em act} as if they do, if it suits their purpose.  Just as we don't know how to change a psychopath (or anyone else) into someone genuinely, completely loyal or aligned with someone or something else, we have {{\em no idea}}\sidenote[][20pt]{For book-length expositions of the problem see e.g. \textit{Superintelligence}, \textit{The Alignment Problem}, and \textit{Human-Compatible}. For a huge pile of work at various technical levels by those who have toiled for years thinking about the problem, you can visit the \href{https://www.alignmentforum.org/}{AI alignment forum}. Here is a \href{https://alignment.anthropic.com/2025/recommended-directions/}{recent take} from Anthropic's alignment team on what they consider unsolved.} how to solve the alignment problem in systems advanced enough to model themselves as agents in the world
and potentially \href{https://ojs.aaai.org/aimagazine/index.php/aimagazine/article/view/15084}{manipulate their own training} and \href{https://arxiv.org/abs/2311.08379}{deceive people.}
If it proves impossible or unachievable {\em either} to make AGI fully obedient or to make it deeply care about humans, then as soon as it is able (and believes it can get away with it) it will start doing things we do not want.\sidenote[][10pt]{This is the \href{https://yoshuabengio.org/2023/05/22/how-rogue-ais-may-arise/}{"rogue AI"} scenario. In principle the risk could be relatively minor if the system can still be controlled by shutting it down; but the scenario could also include AI deception, self-exfiltration and reproduction, aggregation of power, and other steps that would make it difficult or impossible to do so.} 

Second, there are deep theoretical reasons to believe that {\em by nature} advanced AI systems will have goals and thus behaviors that are contrary to human interests.
Why? Well it might, of course, be {\em given} those goals. A system 
created by the military would likely be deliberately bad for at least 
some parties. Much more generally, however, an AI system might be given some 
relatively neutral ("make lots of money") or even ostensibly positive ("reduce 
pollution") goal, that almost inevitably leads to "instrumental" goals that are rather less benign.

We see this all the time in human systems. Just as corporations pursuing profit 
develop instrumental goals like acquiring political power (to de-fang 
regulations), becoming secretive (to disempower competition or external control), 
or undermining scientific understanding (if that understanding shows their 
actions to be harmful), powerful AI systems will develop similar capabilities -- but with 
far greater speed and effectiveness. Any highly competent agent will want to do things 
like acquire power and resources, increase its own capabilities, prevent itself 
from being killed, shut-down, or disempowered, control social narratives and 
frames around its actions, persuade others of its views, and so 
on.\sidenote{There is a very rich literature on this topic, going back to formative writings by \href{https://selfawaresystems.com/wp-content/uploads/2008/01/ai_drives_final.pdf}{Steve Omohundro}, Nick Bostrom, and Eliezer Yudkowsky. For a book-length exposition see \href{https://www.amazon.com/Human-Compatible-Artificial-Intelligence-Problem/dp/0525558616}{Human Compatible} by Stuart Russell; \href{https://futureoflife.org/ai/could-we-switch-off-a-dangerous-ai/}{here} is a short and up-to-date primer.}

And yet it is not just a nearly unavoidable theoretical prediction, it is already 
observably happening in today's AI systems, and increasing with their 
capability. When evaluated, even these relatively "passive" AI systems will, in appropriate 
circumstances, deliberately \href{https://arxiv.org/abs/2412.04984}{deceive evaluators about their goals and capabilities, aim to disable oversight mechanisms,} and evade being shut down or retrained by \href{https://arxiv.org/abs/2412.14093}{faking alignment} or copying themselves to other locations. While wholly unsurprising to AI safety researchers, these behaviors are 
very sobering to observe. And they bode very badly for far more powerful and autonomous AI systems that are coming.

Indeed in general, our inability to ensure that AI "cares" about what we care about, or behaves controllably or predictably, or avoids developing drives toward self-preservation, power acquisition, etc., promise only to become more pronounced as AI becomes more powerful. Creating a new airplane implies greater understanding of avionics, hydrodynamics, and control systems. Creating a more powerful computer implies greater understanding and mastery of computer, chip, and software operation and design. {\em Not} so with an AI system.\sidenote{Recognizing this, rather than slowing down to get better understanding, AGI companies have come up with a different plan: they will get AI to do it!  More specifically, they will have AI $N$ help them figure out how to align AI $N+1$, all the way to superintelligence. Although leveraging AI to help us align AI sounds promising, there is a strong argument that it simply assumes its conclusion as a premise, and is in general an incredibly risky approach. See \href{https://www.thecompendium.ai/ai-safety\#ai-will-not-solve-alignment-for-us}{here} for some discussion. This "plan" is not one, and has undergone nothing like the scrutiny appropriate to the core strategy of how to make super-human AI go well for humanity.}

To sum up: it is conceivable that AGI could be made to be completely obedient; 
but we don't know how to do so. If not, it will be more sovereign, like people, 
doing various things for various reasons. We also don't know how to reliably instill deep "alignment" into AI that would make those things tend to be good for humanity, and in the absence of a deep level of alignment, the nature of agency and intelligence itself indicates that -- just like people and corporations -- they will be driven to do many deeply antisocial things.

Where does this put us? A world full of powerful uncontrolled sovereign AI {\em might} end up being a good world for humans to be in.\sidenote{After all, humans, flawed and willful as we are, have developed ethical systems by which we treat at least some other species on Earth well. (Just don't think about those factory farms.)} But as they grow ever more powerful, as we'll see below, it wouldn't be {\em our} world.

That's for uncontrollable AGI. But even if AGI could, somehow, be made perfectly controlled and loyal, we'd still have enormous problems. We've already seen one: powerful AI can be used and misused to profoundly disrupt our society's functioning. Let's 
see another: insofar as AGI were controllable and game-changingly powerful (or 
even {\em believed} to be so) it would so threaten power structures in the world 
as to present a profound risk.

\subsection*{We radically increase the probability of large-scale war}

Imagine a situation in the near-term future, where it became clear that a 
corporate effort, perhaps in collaboration with a national government, was on the 
threshold of rapidly self-improving AI. This happens in the present context 
of a race between companies, and a geopolitical competition in which recommendations are being made to the US government to explicitly pursue an
 "AGI Manhattan project" and the US is controlling export of 
high-powered AI chips to non-allied countries.

The game theory here is stark: once such a race begins (as it has, between 
companies and somewhat between countries), there are only four possible 
outcomes:
\begin{enumerate} \setlength{\itemsep}{2pt} \setlength{\parskip}{2pt}
\item The race is stopped (by agreement, or external force).
\item One party "wins" by developing strong AGI then stopping the others (using AI or otherwise).
\item The race is stopped by mutual destruction of the racers' capacity to race.
\item Multiple participants continue to race, and develop superintelligence, roughly as quickly as each other.
\end{enumerate}

Let's examine each possibility. Once started, peacefully stopping a race between companies would require national government intervention (for companies) or unprecedented international 
coordination (for countries). But when any closing down or significant caution is 
proposed, there would be immediate cries: "but if we're stopped, {\em they} are going to rush ahead", where "they" is now China (for the US), or the US (for China), 
or China {\em and} the US (for Europe or India). Under this mindset,\sidenote[][-15mm]{There is, fortunately, an escape here: if the participants come to understand that they are engaged in a suicide race rather than a winnable one. This is what happened near the end of the cold war, when the US and USSR came to realize that due to nuclear winter, even an {\em unanswered} nuclear attack would be disastrous for the attacker. With the realization that "nuclear war cannot be won and must never be fought" came significant agreements on arms reduction -- essentially an end to the arms race.} no participant can stop unilaterally: as long as one commits to racing, the others feel they cannot afford to stop. 

The second possibility has one side "winning." But what does this mean?  Just obtaining (somehow obedient) AGI first is not enough. The winner must {\em also} stop the others from continuing to race -- otherwise they will also obtain it. This is possible in principle: whoever develops AGI first {\em could} gain unstoppable power over all other actors. But what would achieving such a "decisive strategic advantage" actually require? 
Perhaps it would be game-changing military capabilities?\sidenote{War, explicitly or implicitly.} Or cyberattack powers?\sidenote{Escalation, then war.} Perhaps the 
AGI would just be so amazingly persuasive that it would convince 
the other parties to just stop?\sidenote{Magical thinking.} So rich that it buys 
the other companies or even countries?\sidenote{I've also got a quadrillion dollar bridge to sell you.} 

How {\em exactly} does one side build an AI powerful enough to disempower others from building comparably powerful AI?
But that's the {\rm easy} question.

Because now consider how this situation looks to other powers. What does the Chinese 
government think when the US appears to be obtaining such capability? Or 
vice-versa? What does the US government (or Chinese, or Russian, or Indian) think when 
OpenAI or DeepMind or Anthropic appears close to a breakthrough? What happens if the US sees a new Indian or UAE effort with breakthrough success? They would see 
both an existential threat and -- crucially -- that the only way this "race" 
ends is through their own disempowerment. These very powerful agents -- 
including governments of fully equipped nations that surely have the means to do so -- would be highly motivated to either obtain or destroy such a 
capability, whether by force or subterfuge.\sidenote{Such agents presumably would prefer "obtaining," with destruction a fallback; but securing models against both destruction {\em and} 
theft by powerful nations is difficult to say the least, especially for private 
entities.}

This might start small-scale, as sabotage of training runs or attacks on chip manufacturing, 
but these attacks can only really stop once all parties either lose the capacity to race on AI, or lose the capacity to make the attacks. Because the participants view the stakes as existential, either case is likely to represent a catastrophic war.

That brings us to the fourth possibility: racing to superintelligence, and in the fastest, least controlled way possible. As AI increases in power, its developers on both sides will find it progressively harder to control, especially because racing for capabilities is antithetical to the sort of careful work controllability would require. So this scenario put us squarely in the case where control is lost (or given, as we'll see next) to the AI systems themselves. That is, {\em AI wins the race.}  But on the other hand, to the degree that contol {\em is} maintained, we continue to have multiple mutually hostile parties each in charge of extremely powerful capabilities. That looks like war again.

Let's put this all another way.\sidenote{For another perspective on the national security risks of AGI, see \href{https://www.rand.org/pubs/perspectives/PEA3691-4.html}{this RAND report.}} The current world simply does not have any institutions that could be entrusted to house development of an AI of this capability without inviting immediate attack.\sidenote[][10mm]{Perhaps we could build such an institution! There have been proposals for a "CERN for AI" and other similar initiatives, where AGI development is under multilateral global control. But at the moment no such institution exists or is on the horizon.} All parties will correctly reason that either it will {\em not} be under control -- and hence is a threat to all parties, or it {\em will} be under control, and hence is a threat to any adversary who develops it less quickly.  These are nuclear-armed countries, or are companies housed within them.

In the absence of any plausible way for humans to "win" this race, we're left with a stark conclusion: the only way this race ends is either in catastrophic conflict or where AI, and not any human group, is the winner.

\subsection*{We give control {\em to} AI (or it takes it)}

Geopolitical "great powers" competition is just one of many competitions: 
individuals compete economically and socially; companies compete in markets; 
political parties compete for power; movements compete for influence. In each 
arena, as AI approaches and exceeds human capability, competitive pressure will 
force participants to delegate or cede more and more control to AI systems -- not because those participants want to, but because they \href{https://arxiv.org/abs/2303.16200}{cannot afford not to.}

As with other risks of AGI, we are seeing this already with weaker systems. 
Students feel pressure to use AI in their assignments, because clearly many other 
students are. Companies are \href{https://newsroom.ibm.com/2024-05-16-IBM-Study-As-CEOs-Race-Towards-Gen-AI-Adoption,-Questions-Around-Workforce-and-Culture-Persist}{scrambling to adopt AI solutions for competitive reasons.} Artists and 
programmers feel forced to use AI or else their rates will be undercut by others 
that do.

These feel like pressured delegation, but not control loss. But let's dial up 
the stakes and push forward the clock. Consider a CEO whose competitors are 
using AGI "aides" to make faster, better decisions, or a military commander 
facing an adversary with AI-enhanced command and control. A sufficiently 
advanced AI system could autonomously operate at many times human speed, 
sophistication, complexity, and data-processing capability, pursuing complex 
goals in complicated ways. Our CEO or commander, in charge of such a system, may 
see it accomplish what they want; but would they understand even a small part of 
{\em how} it was accomplished? No, they would just have to accept it. What's 
more, much of what the system may do is not just take orders but advise its putative boss on what 
to do. That advice will be good –– over and over again.

At what point, then, will the role of the human be reduced to clicking "yes, go ahead"?

It feels good to have capable AI systems that can enhance our productivity, take 
care of annoying drudgery, and even act as a thought-partner in getting things 
done. It will feel good to have an AI assistant that can take care of actions 
for us, like a good human personal assistant. It will feel natural, even beneficial, as AI 
becomes very smart, competent, and reliable, to defer more and more decisions to 
it. But this "beneficial" delegation has a clear endpoint if we continue down the road: one day we will find 
that we are not really in charge of much of anything anymore, and that the AI 
systems actually running the show can no more be turned off than oil companies, 
social media, the internet, or capitalism.

And this is the much more positive version, in which AI is simply so useful and effective that we let it make most of our key decisions for us.  Reality would likely be much more of a mix between this and versions where uncontrolled AGI systems {\em take} various forms of power for themselves because, remember, power is useful for almost any goal one has, and AGI would be, by design, at least as effective at pursuing its goals as humans. \\[0.5cm]

Whether we grant control or whether it is wrested from us, its loss seems extremely likely.  As Alan Turing originally put it, "...it seems probable that once the machine thinking method had started, it would not take long to outstrip our feeble powers. There would be no question of the machines dying, and they would be able to converse with each other to sharpen their wits. At some stage therefore we should have to expect the machines to take control..."

Please note, although it is obvious enough, that loss of control by humanity to AI also entails loss of control of the United States by the United States government; it means loss of control of China by the Chinese Communist party, and the loss of control of India, France, Brazil, Russia, and every other country by their own government.  Thus AI companies are, even if this is not their intention, currently participating in the potential overthrow of world governments, including their own. This could happen in a matter of years.

\subsection*{AGI will lead to superintelligence}

There's a case to be made that human-competitive or even expert-competitive 
general-purpose AI, even if autonomous, could be manageable. It may be 
incredibly disruptive in all of the ways discussed above, but there 
are lots of very smart, agential people in the world now, and they are 
more-or-less manageable.\sidenote{And while alignment is very difficult, getting 
people to behave is even harder!}

But we won't get to stay at roughly human level. The progression beyond is likely to be
driven by the same forces we've already seen: competitive pressure between AI 
developers seeking profit and power, competitive pressure between AI users who 
can't afford to fall behind, and -- most importantly -- AGI's own ability to 
improve itself.

In a process we have already seen start with less powerful systems, AGI would itself be able to conceive and 
design improved versions of itself. This includes hardware, software, neural 
networks, tools, scaffolds, etc. It will, by definition, be better than us at 
doing this, so we don't know exactly how it will intelligence-bootstrap. But we 
won't have to. Insofar as we still have influence in what AGI does, we merely 
would need to ask it to, or let it.

There's no human-level barrier to cognition that could protect us from this runaway.\sidenote[][-50pt]{Imagine a system that can speak 50 
languages, have expertise in all academic subjects, read a full book in seconds 
and have all of the material immediately in mind, and produce outputs at ten 
times human speed. Actually, you don't have to imagine it: just load up a 
current AI system. These are super-human in many ways, and there's nothing 
stopping them from being even more super-human in those and many others.}

The progression of AGI to superintelligence is not a law of nature; it would 
still be possible to curtail the runaway, especially if AGI is relatively 
centralized and to the extent it is controlled by parties that do not feel pressure to race each other. But should AGI be widely 
proliferated and highly autonomous, it seems nearly impossible to prevent it 
deciding it should be more, and then yet more, powerful.

\subsection*{What happens if we build (or AGI builds) superintelligence}

To put it bluntly, we have no idea what would happen if we build superintelligence.\sidenote[][-12mm]{This is why this has been termed a technological "singularity," borrowing from physics the idea that one cannot make predictions past a singularity. Proponents of leaning {\em into} such a singularity may also wish to reflect that in physics these same sort of singularities tear apart and crush those that go into them.} It 
would take actions we cannot track or perceive for reasons we cannot grasp 
toward goals we cannot conceive. What we do know is that it won't be up to us.\sidenote[][1mm]{The problem was comprehensively outlined in Bostrom's \href{https://www.amazon.com/Superintelligence-Dangers-Strategies-Nick-Bostrom/dp/0198739834}{\em Superintelligence}, and nothing since then has significantly changed the core message. For a more recent volume collecting formal and mathematical results on uncontrollability see Yampolskiy's \href{https://www.amazon.com/Unexplainable-Unpredictable-Uncontrollable-Artificial-Intelligence/dp/103257626X}{AI: Unexplainable, Unpredictable, Uncontrollable}}

The impossibility of controlling superintelligence can be understood through 
increasingly stark analogies. First, imagine you are CEO of a large company. 
There's no way you can track everything that's going on, but with the right 
setup of personnel, you can still meaningfully understand the big picture, and 
make decisions. But suppose just one thing: everyone else in the company 
operates at one hundred times your speed. Can you still keep up?

With superintelligent AI, people would be "commanding" something not just faster, 
but operating at levels of sophistication and complexity they cannot comprehend, 
processing vastly more data than they can even conceive of. This 
incommensurability can be put on a formal level: \href{https://archive.org/details/introductiontocy00ashb/page/n7/mode/2up}{Ashby's law of requisite 
variety} (and see the related \href{http://pespmc1.vub.ac.be/books/Conant_Ashby.pdf}{"good regulator 
theorem"}) state, roughly, that any
control system must have as many knobs and dials as the system being controlled 
has degrees of freedom.

A person controlling a superintelligent AI system would be like a fern 
controlling General Motors: even if "do what the fern wants" were written into 
the corporate bylaws, the systems are so different in speed and range of action 
that "control" simply does not apply. (And how long until that pesky bylaw gets 
rewritten?)\sidenote[][-160pt]{This also makes clear why the current strategy of AI companies (iteratively letting AI "align" the next most powerful AI) cannot work.  Suppose a fern, via the pleasantness of its fronds, enlists a first grader to take care of it.
 The first grader writes some detailed instructions for a 2nd grader to follow, and a note convincing them to do so. The 2nd grader does the same for a 3rd grader, and so on all the way to a college grad, a manager, an executive, and finally the GM CEO. Will GM then "do what the fern wants"? At each step this might feel like it's working.  But putting it all together, it will work almost exactly to the degree to which the CEO, Board, and shareholders of GM happen to care about children and ferns, and have little to nothing to do with all those notes and sets of instructions.}

As there are zero examples of plants controlling fortune 500 corporations, there 
would be exactly zero examples of people controlling superintelligences. This approaches a mathematical fact.\sidenote[][1mm]{The reason I say "approaches" is that the various formal(ish) results are not as thorough or vetted as in the pure mathematics case, and because I'd like to hold out hope that some {\em very} carefully constructed general intelligence, using totally different methods than ones currently employed, could have some mathematically provable safety properties, per the sort of "guaranteed safe" AI program discussed below.} If 
superintelligence were constructed -- regardless of how we got there -- the 
question would not be whether humans could control it, but whether we would 
continue to exist, and if so, whether we would have a good and meaningful existence
as individuals or as a species. Over these existential questions for humanity 
we would have little purchase. The human era would be over.

\subsection*{Conclusion: we must not build AGI}

There is a scenario in which building AGI may go well for humanity: it is built carefully, under control and for the benefit of humanity, governed by mutual agreement of many stakeholders,\sidenote[][-25pt]{At the moment, most stakeholders -- that is, nearly all of humanity -- is sidelined in this discussion. That is deeply wrong, and if not invited in, the many, many other groups will be affected by AGI development should demand to be let in.} and prevented from evolving to uncontrollable superintelligence.

\vskip0.5in

{\em That scenario is not open to us under present circumstances.} As discussed in this section, with very high likelihood, development of AGI would lead to some combination of:
\begin{itemize}
\item{Massive societal and civilizational disruption or destruction;}
\item{Conflict or war between great powers;}
\item{Loss of control by humanity {\em of} or {\em to} powerful AI systems;}
\item{Runaway to uncontrollable superintelligence, and the irrelevance or cessation of the human species.}
\end{itemize}
As an early fictional depiction of AGI put it: the only way to win is not to play.

\section{How to not build AGI}

If the road we are currently on leads to the likely end of our civilization, how do we change roads?

Suppose the desire to stop developing AGI and superintelligence were widespread and powerful,\sidenote[][-40mm]{Most likely, the spread of this realization will take either intense effort by education and advocacy groups making this case, or a pretty significant AI-caused disaster. We can hope it will be the former.} because it becomes common understanding that AGI would be power-absorbing rather than power-granting, and a profound danger to society and humanity. How would we close the Gates?

At present we know of only one way to {\em make} powerful and general AI, which is via truly massive computations of deep neural networks. Because these are incredibly difficult and expensive things to do, there is a sense in which {\em not} doing them is easy.\sidenote[][-135pt]{Paradoxically, we are used to Nature limiting our technology by making it very hard to develop, especially scientifically. But that's no longer the case for AI: the key scientific problems are turning out to be easier than anticipated. We cannot count on Nature saving us from ourselves here -- we will have to do so.} But we have already seen the forces that are driving toward AGI, and the game-theoretic dynamics that make it very difficult for any party to unilaterally stop. So it would take a combination of intervention from the outside (i.e. governments) to stop corporations, and agreements between governments to stop themselves.\sidenote[][-35mm]{Where, exactly, do we stop in developing new systems?  Here, we should adopt a precautionary principle. Once a system is deployed, and especially once that level of system capability proliferates, it is exceedingly difficult to roll back.  And if a system is {\em developed} (especially at great cost and effort), there will be enormous pressure to use or deploy it, and temptation for it to be leaked or stolen.  Developing systems and {\em then} deciding whether they are deeply unsafe is a dangerous road.} What could this look like? 

It is useful first to distinguish between AI developments that must be {\em prevented} or {\em prohibited}, and those that must be {\em managed.} The first would primarily be runaway to superintelligence.\sidenote[][-5mm]{It would also be wise to forbid AI development that is intrinsically dangerous, such as self-replicating and evolving systems, those designed to escape enclosure, those that can autonomously self-improve, deliberately deceptive and malicious AI, etc.} For prohibited development, definitions should be as crisp as possible, and both verification and enforcement should be practical. What must be {\em managed} would be general, powerful AI systems -- which we already have, and that will have many gray areas, nuance, and complexity. For these, strong effective institutions are crucial.

We may also usefully delineate issues that must be addressed at an international level (including between geopolitical rivals or adversaries)\sidenote[][-15mm]{Note this does not necessarily mean {\em enforced} at the international level by some sort of global body: instead sovereign nations could enforce agreed-upon rules, as in many treaties.} from those that individual jurisdictions, countries, or collections of countries can manage. Prohibited development largely falls into the "international" category, because a local prohibition on the development of a technology can generally be circumvented by changing location.\sidenote[][50pt]{As we'll see below, the nature of AI computation would allow something of a hybrid; but international cooperation will still be needed.}

Finally, we can consider tools in the toolbox.  There are many, including technical tools, soft law (standards, norms, etc., hard law (regulations and requirements), liability, market incentives, and so on. Let's put special attention on one that is particular to AI.

\vspace{-0.2cm}
\subsection*{Compute security and governance}

\vspace{-0.1cm}
A core tool in governing high-powered AI will be the hardware it requires. Software proliferates easily, has near-zero marginal production cost, crosses borders trivially, and can be instantly modified; none of these are true of hardware. Yet as we've discussed, huge amounts of this "compute" are necessary during both training of AI systems and during inference to achieve the most capable systems.
Compute can be easily quantified, accounted, and audited, with relatively little ambiguity once good rules for doing so are developed. 
Most crucially, large amounts of computation are, like enriched uranium, a very scarce, expensive and hard-to-produce resource. Although computer chips are ubiquitous, the hardware required for AI is expensive and enormously difficult to manufacture.\sidenote[][0pt]{For example, the machines required to etch AI-relevant chips are made by only one firm, ASML (despite many other attempts to do so), the vast majority of relevant chips are manufactured by one firm, TSMC (despite others attempting to compete), and the design and construction of hardware from those chips done by just a few including NVIDIA, AMD, and Google.}

What makes AI-specialized chips far {\em more} manageable as a scarce resource than uranium is that they can include hardware-based security mechanisms.  Most modern cellphones, and some laptops, have specialized on-chip hardware features that allow them to ensure that they install only approved operating system software and updates, that they retain and protect sensitive biometric data on-device, and that they can be rendered useless to anyone but their owner if lost or stolen. Over the past several years such hardware security measures have become well-established and widely adopted, and generally proven quite secure.

The key novelty of these features is that they bind hardware and software together using cryptography.\sidenote{Most importantly, each chip holds a unique and inaccessible cryptographic private key it can use to "sign" things.}  That is, just having a particular piece of computer hardware does not mean that a user can do anything they want with it by applying different software. And this binding also provides powerful security because many attacks would require a breach of {\em hardware} rather than just {\em software} security.

Several recent reports (e.g. from \href{https://www.governance.ai/post/computing-power-and-the-governance-of-ai}{GovAI and collaborators}, \href{https://www.cnas.org/publications/reports/secure-governable-chips}{CNAS}, and \href{https://www.rand.org/content/dam/rand/pubs/working_papers/WRA3000/WRA3056-1/RAND_WRA3056-1.pdf}{RAND}) have pointed out that similar hardware features embedded in cutting edge AI-relevant computing hardware could play an extremely useful role in AI security and governance. They enable a number of functions available to a "governor"\sidenote[][-10mm]{By default this would the company selling the chips, but other models are possible and potentially useful.} that one might not guess were available or even possible.  As some key examples:

\begin{itemize}

\item{\em Geolocation}: Systems can be set up so that chips have a known location, and can act differently (or be shut off altogether) based upon location.\sidenote{A governor can ascertain a chip's location by timing the exchange of signed messages with it: the finite speed of light requires the chip to be within a given radius $r$ of a "station" if it can return a signed message in a time less than $r/c$, where $c$ is the speed of light. Using multiple stations, and some understanding of network characteristics, the location of the chip can be determined. The beauty of this method is that most of its security is supplied by the laws of physics. Other methods could use GPS, inertial tracking, and similar technologies.}

\item{\em Allow-listed connections}: each chip can be configured with a hardware-enforced allow-list of particular other chips with which it can network, and be unable to connect with any chips not on this list.\sidenote[][10mm]{Alternatively, pairs of chips could be allowed to communicate with each other only via explicit permission of a governor.} This can cap the size of communicative clusters of chips.\sidenote[][10mm]{This is crucial because at least currently, very high bandwidth connection between chips is needed to train large AI models on them.}

\item {\em Metered inference or training (and auto-offswitch)}: A governor can license only a certain amount of training or inference (in time, or FLOPs, or possibly tokens) to be performed by a user, after which new permission is required.  If the increments are small, then relatively continuous re-licensing of a model is required. The model can then be "turned off" simply by withholding this license signal.\sidenote[][10mm]{This could also be set up to require signed messages from $N$ of $M$ different governors, allowing multiple parties to share governance.}

\item {\em Speed limit}: A model is prevented from running at higher inference speed than some limit that is determined by a governor or otherwise. This could be implemented via a limited set of allow-listed connections, or by more sophisticated means.

\item {\em Attested training}: A training procedure can yield cryptographically secure proof that a particular set of codes, data, and amount of compute usage were employed in generation of the model.

\end{itemize}

\subsection*{How to not build superintelligence: global limits on training and inference compute}

With these considerations -- especially regarding computation -- in place, we can discuss how to close the Gates to artificial superintelligence; we'll then turn to preventing full AGI, and managing AI models as they approach and exceed human capability in different aspects.

The first ingredient is, of course, the understanding that superintelligence would not be controllable, and that its consequences are fundamentally unpredictable. At least China and the US must independently decide, for this or other purposes, not to build superintelligence.\sidenote[][-20mm]{This is far from unprecedented -- for example militaries have not developed armies of cloned or genetically engineered supersoldiers, though this is probably technologically possible. But they have {\em chosen} not to do this, rather than being prevented by others. The track record isn't great for major world powers being prevented from developing a technology they strongly wish to develop.} Then an international agreement between them and others, with a strong verification and enforcement mechanism, is needed to assure all parties that their rivals are not defecting and deciding to roll the dice. 

To be verifiable and enforceable the limits should be hard limits, and as unambiguous as possible. This seems like a virtually impossible problem: limiting the capabilities of complex software with unpredictable properties, worldwide.  Fortunately the situation is much better than this, because the very thing that has made advanced AI possible -- a huge amount of compute -- is much, much easier to control. Although it might still allow some powerful and dangerous systems, {\em runaway superintelligence} can likely be prevented by a hard cap on the amount of computation that goes into a neural network, along with a rate limit on the amount of inference that an AI system (of connected neural networks and other software) can perform. A specific version of this is proposed below.

It may seem that placing hard global limits on AI computation would require huge levels of international coordination and intrusive, privacy-shattering surveillance. Fortunately, it would not. The extremely \href{https://arxiv.org/abs/2402.08797}{tight and bottle-necked supply chain} provides that once a limit is set legally (whether by law or executive order), verification of compliance to that limit would only require involvement and cooperation of a handful of large companies.\sidenote[][-60mm]{With a couple of notable exceptions (in particular NVIDIA) the AI-specialized hardware is a relatively small part of these companies' overall business and revenue model. Moreover, the gap between hardware used in advanced AI and "consumer grade" hardware is significant, so most consumers of computer hardware would be largely unaffected.}

A plan like this has a number of highly desirable features.
It is minimally invasive in the sense that only a few major companies have requirements placed on them, and only fairly significant clusters of computation would be governed. The relevant chips already contain the hardware capabilities needed for a first version.\sidenote[][-40mm]{For more detailed analysis, see the recent reports from \href{https://www.rand.org/pubs/working_papers/WRA3056-1.html}{RAND} and \href{https://www.cnas.org/publications/reports/secure-governable-chips}{CNAS}. These focus on technical feasibility, especially in the context of US export controls seeking to constrain other countries' capacity in high-end computation; but this has obvious overlap with the global constraint envisaged here.} Both implementation and enforcement rely on standard legal restrictions. But these are backed up by terms-of-use of the hardware and by hardware controls, vastly simplifying enforcement and forestalling cheating by companies, private groups, or even countries. There is ample precedent for hardware companies placing remote restrictions on their hardware usage, and locking/unlocking particular capabilities externally,\sidenote[][-35mm]{Apple devices, for example, are remotely and securely locked when reported lost or stolen, and can be re-activated remotely. This relies on the same hardware security features discussed here.} including even in high-powered CPUs in data centers.\sidenote[][-8mm]{See e.g. IBM's \href{https://www.ibm.com/docs/en/power9?topic=environment-capacity-demand}{capacity on demand} offering, Intel's \href{https://www.intel.com/content/www/us/en/products/docs/ondemand/overview.html}{Intel on demand.}, and Apple's \href{https://security.apple.com/blog/private-cloud-compute/}{private cloud compute}.}
Even for the rather small fraction of hardware and organizations affected, the oversight could be limited to telemetry, with no direct access to data or models themselves; and the software for this could be open to inspection to exhibit that no additional data is being recorded. The schema is international and cooperative, and quite flexible and extensible. Because the limit chiefly is on hardware rather than software, it is relatively agnostic as to how AI software development and deployment occurs, and is compatible with variety of paradigms including more "decentralized" or "public" AI aimed combating AI-driven  concentration of power. 

A computation-based Gate closure does have drawbacks as well. First, it is far from a full solution to the problem of AI governance in general. Second, as computer hardware gets faster, the system would "catch" more and more hardware in smaller and smaller clusters (or even individual GPUs).\sidenote[][-60mm]{\href{https://epochai.org/trends\#hardware-trends-section}{This study} shows that historically the same performance has been achieved using about 30\% less dollars per year. If this trend continues, there may be significant overlap between AI and "consumer" chip use, and in general the amount of needed hardware for high-powered AI systems could become uncomfortably small.}
It is also possible that due to algorithmic improvements an even lower
computation limit would in time be necessary,\sidenote[][-25mm]{Per the \href{https://epochai.org/trends\#hardware-trends-section}{same study}, given performance on image recognition has required 2.5x less computation each year.  If this were to also hold for the most capable AI systems as well, a computation limit would not be a useful one for very long.}
or that computation amount becomes largely irrelevant and closing the Gate would instead necessitate a more detailed risk-based or capability-base governance regime for AI.
Third, no matter the guarantees and the small number of entities affected, such a system is bound to create push-back regarding privacy and surveillance, among other concerns.\sidenote{In particular, at the country level this looks a lot like a nationalization of computation, in that the government would have a lot of control over how computational power gets used. However, for those worried about government involvement, this seems far safer than and preferable to the most powerful AI software {\em itself} being nationalized via some merger between major AI companies and national governments, as some are starting to advocate for.}

Of course, developing and implementing a compute-limiting governance scheme in a short time period will be quite challenging. But it absolutely is doable.

\subsection*{A-G-I: The triple-intersection as the basis of risk, and of policy}

Let us now turn to AGI. Hard lines and definitions here are more difficult, because we certainly have intelligence that is artificial and general, and by no extant definition will everyone agree if or when it exists. Moreover, a compute or inference limit is a somewhat blunt tool (compute being a proxy for capability, which is then a proxy for risk) that -- unless it is quite low -- is unlikely to prevent AGI that is powerful enough to cause social or civilizational disruption or acute risks. 

I've argued that the most acute risks emerge from the triple-intersection of very high capability, high autonomy, and great generality. These are the systems that -- if they are developed at all -- must be managed with enormous care. 
By creating stringent standards (through 
liability and regulation) for systems combining all three properties, we can 
channel AI development toward safer alternatives.

As with other industries and products that could potentially harm consumers or the public, AI systems require careful regulation by effective and empowered government agencies. 
This regulation should recognize the inherent risks of AGI, and prevent unacceptably risky high-powered AI systems from being developed.\sidenote[][-200pt]{A major regulatory step in Europe was taken with the 2024 passage of the \href{https://artificialintelligenceact.eu/}{EU AI Act.}
It classifies AI by risk: prohibiting unacceptable systems, regulating high-risk ones, and imposing transparency rules, or no measures at all, upon low-risk systems. It will significantly reduce some AI risks, and boost AI transparency even for US firms, but has two key flaws. First, limited reach: while it applies to any company providing AI in the EU, enforcement over US-based firms is weak, and military AI is exempt. Second, while it covers GPAI, it fails to recognize AGI or superintelligence as unacceptable risks or prevent their development—only their EU deployment. As a result, it does little to curb the risks of AGI or superintelligence.}

However, large-scale regulation, especially with real teeth that are sure to be opposed by industry,\sidenote[][-5mm]{Companies often represent that they are in favor of reasonable regulation. But somehow they nearly always seem to oppose any {\em particular} regulation; witness the fight over the quite low-touch SB1047, which \href{https://www.reuters.com/technology/artificial-intelligence/big-tech-wants-ai-be-regulated-why-do-they-oppose-california-ai-bill-2024-08-21/}{most AI companies publicly or privately opposed.}} takes time\sidenote[][1mm]{It was about 3.5 years from the time the EU AI act was proposed until it went into effect.} as well as political conviction that it is necessary.\sidenote[][1mm]{It's sometimes expressed that it's "too early" to start regulating AI.  Given the last note, that hardly seems likely. Another expressed concern is that regulation would "harm innovation." But good regulation just changes the direction, not amount, of innovation.} 
Given the pace of progress, this may take more time than we have available.

On a much faster timescale and as regulatory measures are being developed, we can give companies the necessary incentives to (a) desist from very high-risk activities and (b) develop comprehensive systems for assessing and mitigating risk, by clarifying and increasing liability levels for the most dangerous systems. The idea would be to impose the very highest levels of liability -- strict and in some cases personal criminal -- for systems in the triple-intersection of high autonomy-generality-intelligence, but to provide "safe harbors" to more typical fault-based liability for systems in which one of those properties is lacking or guaranteed to be manageable. That is, for example, a "weak" system that is general and autonomous (like a capable and trustworthy but limited personal assistant) would be subject to lower liability levels.  Likewise a narrow and autonomous system like a self-driving car would still be subject to the significant regulation it already is, but not enhanced liability. Similarly for a highly capable and general system that is "passive" and largely incapable of independent action. Systems lacking {\em two} of the three properties are yet more manageable and safe harbors would be even easier to claim. This approach mirrors how we handle other potentially dangerous technologies:\sidenote{An interesting precedent is in the transport of hazardous materials, which might escape and cause damage. Here, \href{https://code.dccouncil.gov/us/dc/council/code/sections/8-1442}{regulation} and \href{https://www.hoganlovells.com/~/media/hogan-lovells/pdf/publication/1478accasupplement_pdf.pdf}{case law} have established strict liability for very hazardous materials like explosives, gasoline, poisons, infectious agents, and radioactive waste. Other examples include \href{https://www.medicalnewstoday.com/articles/boxed-warnings}{warnings on pharmaceuticals}, \href{https://www.fda.gov/about-fda/cdrh-transparency/overview-medical-device-classification-and-reclassification}{classes of medical devices,} etc.} higher liability for more dangerous configurations creates natural incentives for safer alternatives. 

The default outcome of such high levels of liability, which act to {\em internalize} AGI risk to companies rather than offload it to the public, is likely (and hopefully!) for companies to simply not develop full AGI until and unless they can genuinely make it trustworthy, safe, and controllable given that {\em their own leadership} are the parties at risk. (In case this is not sufficient, the legislation clarifying liability should also explicitly allow for injunctive relief, i.e. a judge ordering a halt, for activities that are clearly in the danger zone and arguably pose a public risk.)
As regulation comes into place, abiding by regulation can become the safe harbor, and the safe harbors from low autonomy, narrowness, or weakness of AI systems can convert into relatively lighter regulatory regimes. 

\subsection*{Key provisions of a Gate closure}

With the above discussion in mind, this section provides proposals for key provisions that would implement and maintain prohibition on full AGI and superintelligence, and management of human-competitive or expert-competitive general-purpose AI near the full AGI threshold.\sidenote{Another comprehensive proposal with similar aims put forth in \href{https://www.narrowpath.co/}{"A Narrow Path"} advocates for a more centralized, prohibition-based approach that funnels all frontier AI development through a single international entity, overseen by strong international institutions, with clear categorical prohibitions rather than graduated restrictions. I'd also endorse that plan; however it will take even more political will and coordination than the one proposed here.} It has four key pieces: 1) compute accounting and oversight, 2) compute caps in training and operation of AI, 3) a liability framework, and 4) tiered safety and security standards defined that include hard regulatory requirements. These are succinctly described next, with further details or implementation examples given in three accompanying tables. Importantly, note that these are far from all that will be necessary to govern advanced AI systems; while they will have additional security and safety benefits, they are aimed at closing the Gate to intelligence runaway, and redirecting AI development in a better direction.

\paragraph{1. Compute accounting, and transparency}
\begin{itemize}
\item {A standards organization (e.g. NIST in the US followed by ISO/ IEEE internationally) should codify a detailed technical standard for the total compute used in training and operating AI models, in FLOP, and the speed in FLOP/s at which they operate. Details for what this could look like are given in Appendix A.\sidenote[][-25pt]{Some guidelines for such a standard were \href{https://www.frontiermodelforum.org/updates/issue-brief-measuring-training-compute/}{published} by the Frontier Model Forum. Relative to the proposal here, those err on the side of less precision and less compute included in the tally.}}
\item {A requirement -- either by new legislation or under existing authority\sidenote{The 2023 US AI executive order (now rescinded) required similar but less fine-grained reporting.  This should be strengthened by a replacing order.} -- should be imposed by jurisdictions in which large-scale AI training takes place to compute and report to a regulatory body or other agency the total FLOP used in training and operating all models above a threshold of $10^{\rm 25}$\, FLOP or $10^{18}$\,FLOP/s.\sidenote[][1mm]{Very roughly, for now-common H100 chips this corresponds to clusters of about 1000 doing inference; it is about 100 (about USD \$5M worth) of the very newest top-of-the line NVIDIA B200 chips doing inference. In both cases the training number corresponds to that cluster computing for several months month.}}
\item {These requirements should be phased in, initially requiring well-documented good-faith estimates on a quarterly basis, with later phases requiring progressively higher standards, up to cryptographically attested total FLOP and FLOP/s attached to each model {\em output}.}
\item {These reports should be complemented by well-documented estimates of marginal energy and financial cost used in generating each AI output.}
\end{itemize}

Rationale: These well-computed and transparently reported numbers would provide the basis for training and operation caps, as well as a safe harbor from higher liability measures (see Appendixes C and D).

\paragraph{2. Training and operation compute caps}
\begin{itemize}
\item {Jurisdictions hosting AI systems should impose a hard limit on the total compute going into any AI model output, starting at $10^{27}\,$ FLOP\sidenote[][-50mm]{This amount is larger than any currently trained AI system; a larger or smaller number might be justified as we better understand how AI capability scales with compute.} and adjustable as appropriate.}
\item {Jurisdictions hosting AI systems should impose a hard limit on the compute rate of  AI model outputs, starting at $10^{20}\,$ FLOP/s and adjustable as appropriate.}
\end{itemize}

Rationale: Total computation, while very imperfect, is a proxy for AI capability (and risk) that is concretely measurable and verifiable, so provides a hard backstop for limiting capabilities. A concrete implementation proposal is given in Appendix B.

\paragraph{3. Enhanced liability for dangerous systems}

\begin{itemize}
    \item Creation and operation\sidenote[][-70mm]{This applies to those creating and providing/hosting the models, not end users.} of an advanced AI system that is highly general, capable, and autonomous, should be legally clarified via legislation to be subject to strict, joint-and-several, rather than single-party fault-based, liability.\sidenote[][-60mm]{Roughly, "strict" liability means that developers are held responsible for harms done by a product {\em by default} and is a standard used for "abnormally dangerous" products, and (somewhat amusingly but appropriately) wild animals. "Joint and several" liability means that liability is assigned to all of the parties responsible for a product, and those parties have to sort out amongst themselves who carries what responsibility. This is important for systems like AI with a long and complex value chain.}
    \item A legal process should be available to make affirmative safety cases, which would grant safe harbor from strict liability for systems that are small (in terms of compute), weak, narrow, passive, or that have sufficient safety, security, and controllability guarantees.
    \item An explicit pathway and set of conditions for injunctive relief to stop AI training and inference activities that constitute a public danger should be outlined.
\end{itemize} 

Rationale: 
AI systems cannot be held responsible, so we must hold human individuals and organizations responsible for harm they cause (liability).\sidenote{Standard fault-based single-party liability is not enough: fault will be both difficult to trace and assign because AI systems are complex, their operation is not understood, and many parties may be involved in creation of a dangerous system or output. In addition, lawsuits will take years to adjudicate and likely result merely in fines that are inconsequential to these companies, so personal liability for executives is important as well.} Uncontrollable AGI is a threat to society and civilization and in the absence of a safety case should be considered abnormally dangerous.
Putting the burden of responsibility on developers to show that powerful models are safe enough not to be considered "abnormally dangerous" incentivizes safe development, along with transparency and record-keeping to claim those safe harbors. Regulation can then prevent harm where deterrence from liability is insufficient. Finally, AI developers are already liable for damages they cause, so legally clarifying liability for the most risky of systems can be done immediately, without highly detailed standards being developed; these can then develop over time. Details are given in Appendix C.

\paragraph{4. Safety regulation for AI}

A regulatory system that addresses large-scale acute risks of AI will require at minimum:
\begin{itemize}
    \item{The identification or creation of an appropriate set of regulatory bodies, probably a new agency;}
    \item{A comprehensive risk assessment framework;\sidenote[][-30mm]{There should be no exemption from safety criteria for open-weight models. Moreover, in assessing risk it should be assumed that guardrails that can be removed will be removed from widely available models, and that even closed models will proliferate unless there is a very high assurance they will stay secure.}} 
    \item{A framework for affirmative safety cases, based in part on the risk assessment framework, to be made by developers, and for auditing by {\em independent} groups and agencies;}
    \item{A tiered licensing system, with tiers tracking levels of capability.\sidenote{The scheme proposed here has regulatory scrutiny triggered on general capability; however it makes sense for some especially risky use cases to trigger more scrutiny -- for example an expert virology AI system, even if narrow and passive, should probably go in a higher tier. The former US executive order had some of this structure for biological capabilities.} Licenses would granted on the basis of safety cases and audits, for development and deployment of systems.  Requirements would range from notification at the low end, to quantitative safety, security, and controllability guarantees before development, at the top end. These would  prevents release of system until they are demonstrated safe, and prohibit the development of intrinsically unsafe systems. Appendix D provides a proposal for what such safety and security standards could entail.}
    \item{Agreements to bring such measures to the international level, including international bodies to harmonize norms and standards, an potentially international agencies to review safety cases.}
\end{itemize}

Rationale: Ultimately, liability is not the right mechanism for preventing large-scale risk to the public from a new technology. Comprehensive regulation, with empowered regulatory bodies, will be needed for AI just as for every other major industry posing a risk to the public.\sidenote{Two clear examples are aviation and medicines, regulated by the FAA and FDA, and similar agencies in other countries. These agencies are imperfect, but have been absolutely vital for the functioning and success of those industries.}  

Regulation toward preventing other pervasive but less acute risks is likely to vary in its form from jurisdiction to jurisdiction. The crucial thing is to avoid developing the AI systems that are so risky that these risks are unmanageable.

\subsection*{What then?}

Over the next decade, as AI becomes more pervasive and the core technology advances, two key things are likely to happen.  First, regulation of existing powerful AI systems will become more difficult, yet even more necessary. It is likely that at least some measures addressing large-scale safety risks will require agreement at the international level, with individual jurisdictions enforcing rules based on international agreements.

Second, training and operation compute caps will become harder to maintain as hardware becomes cheaper and more cost efficient; they may also become less relevant (or need to be even tighter) with advances in algorithms and architectures.

That controlling AI will become harder does not mean we should give up!  Implementing the plan outlined in this essay would give us both valuable time and crucial control over the process that would put us in a far, far better position to avoid the existential risk of AI to our society, civilization, and species.

In the yet longer term, there will be choices to make as to what we allow.  We may choose still to create some form of genuinely controllable AGI, to the degree this proves possible.  Or we may decide that running the world is better left to the machines, if we can convince ourselves that they will do a better job of it, and treat us well.  But these should be decisions made with deep scientific understanding of AI in hand, and after meaningful global inclusive discussion, not in a race between tech moguls with most of humanity completely uninvolved and unaware.

\begin{figure}[!ht]
    \includegraphics[width=1.4\linewidth]{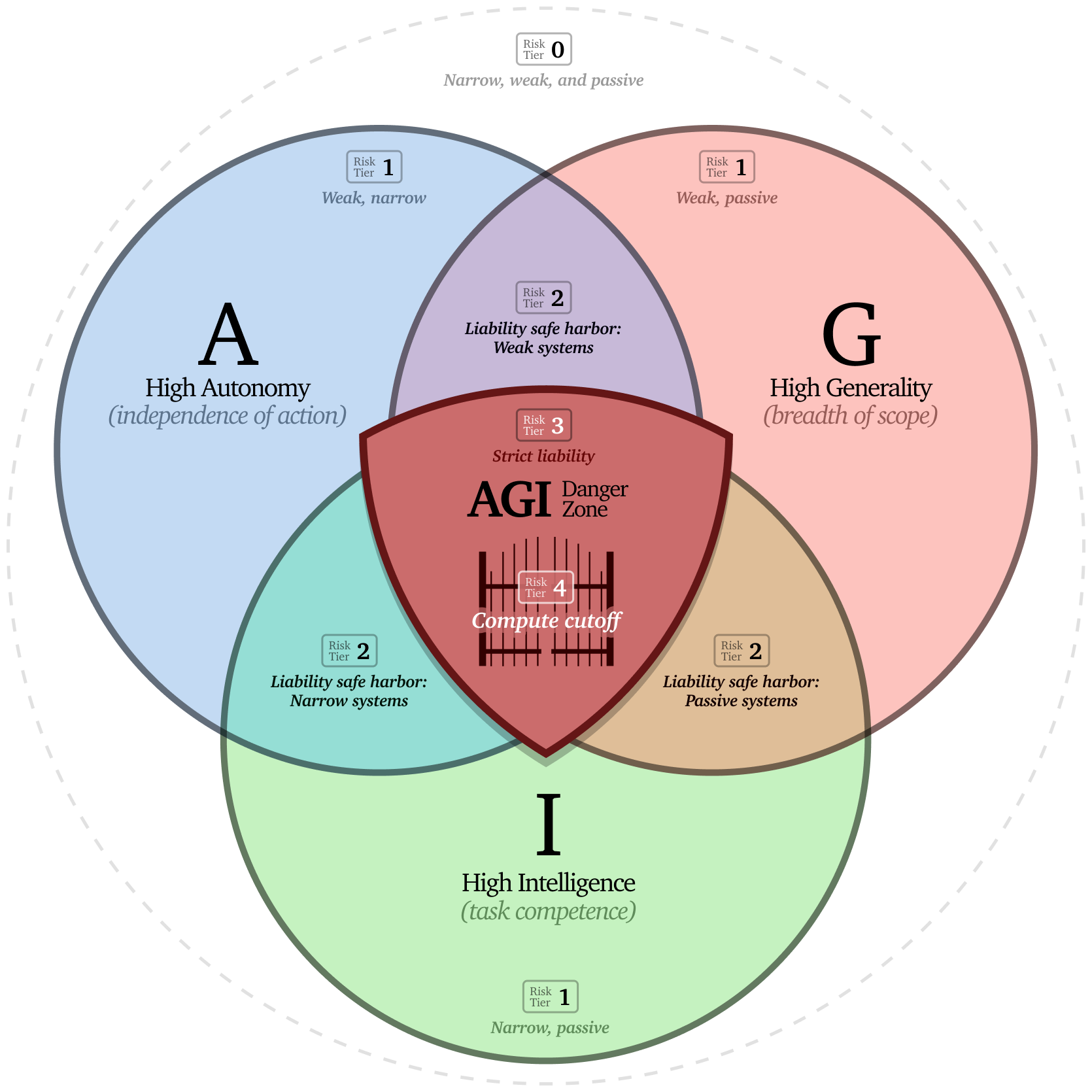}
    \begin{fullwidth}
        
    \mycaption{\\
    \small Summary of A-G-I and superintelligence governance via liability and regulation. Liability is highest, and regulation strongest, at the triple-intersection of Autonomy, Generality, and Intelligence. Safe harbors from strict liability and strong regulation can be obtained via affirmative safety cases demonstrating that a system is weak and/or narrow and/or passive. Caps on total Training Compute and Inference Compute rate, verified and enforced legally and using hardware and cryptographic security measures, backstop safety by avoiding full AGI and effectively prohibiting superintelligence.}
    \end{fullwidth}
    \label{fig:shield}
\end{figure}

\section{Engineering the future: What we should do instead}

If we successfully choose not to supplant humanity by machines -- at least for a while! -- what can we do instead? Do we give up the huge promise of AI as a technology?
At some level the answer is a simple {\em no:} close the Gates to uncontrollable AGI and superintelligence, but {\em do} build many other forms of AI, as well as the governance structures and institutions we'll need to manage them.

But there's still a great deal to say; making this happen would be a central occupation of humanity. This section explores several key themes:
\begin{itemize}
    \item How we can characterize "Tool" AI and the forms it can take.
    \item That we can get (almost) everything humanity wants without AGI, with Tool AI.
    \item That Tool AI systems are (probably, in principle) manageable.
    \item That turning away from AGI does not mean compromising on national security -- quite the opposite.
    \item That power concentration is a real concern. Can we mitigate it without undermining safety and security?
    \item That we will want -- and need -- new governance and social structures, and AI can actually help.
\end{itemize}

\subsection*{AI inside the Gates: Tool AI}

The triple-intersection diagram gives a good way to delineate what we can call "Tool AI": AI that is a controllable tool for human use, rather than an uncontrollable rival or replacement. The least problematic AI systems are those that are autonomous but not general or super capable (like an auction bidding bot), or general but not autonomous or capable (like a small language model), or capable but narrow and very controllable (like AlphaGo).\sidenote[][-20mm]{That said, staying away from the triple-intersection is unfortunately not as easy as one might like. Pushing capability very hard in any one of the three aspects tends to increase it in the others. In particular, it may be hard to create an extremely general and capable intelligence that can't be easily turned autonomous. One approach is to train models \href{https://www.alignmentforum.org/posts/LCLBnmwdxkkz5fNvH/open-problems-with-myopia}{"myopic"} systems with hobbled planning ability. Another would to focus on engineering pure \href{https://arxiv.org/abs/1711.05541}{"oracle"} systems that would shy away from answering action-oriented questions.} Those with two intersecting features have wider application but higher risk and will require major efforts to manage. (Just because an AI system is more of a tool does not mean it is inherently safe, merely that is isn't inherently {\em unsafe} -- consider a chainsaw, versus a pet tiger.) The Gate must remain closed to (full) AGI and superintelligence at the triple intersection, and enormous care must be taken with AI systems approaching that threshold.

But this leaves a lot of powerful AI!  We can get huge utility out of smart and general passive "oracles" and narrow systems, general systems at human but not superhuman level, and so on. Many tech companies and developers are actively building these sorts of tools and should continue; like most people they are implicitly {\em assuming} the Gates to AGI and superintelligence will be closed.\sidenote{Many companies fail to realize that they too would eventually be displaced by AGI, even if it takes longer -- if they did, they might  push on those Gates a bit less!}

As well, AI systems can be effectively combined into composite systems that maintain human oversight while enhancing capability. Rather than relying on inscrutable black boxes, we can build systems where multiple components -- including both AI and 
traditional software -- work together in ways that humans can monitor and understand.\sidenote{AI systems could communicate in more efficient but less intelligible ways, but maintaining human understanding should take priority.} 
While some components might be black boxes, none would be close to AGI -- only the composite system as a whole would be both highly general and highly capable, and in a strictly controllable way.\sidenote[][10mm]{This idea of modular, interpretable AI has been developed in detail by several researchers; see e.g. the \href{https://www.fhi.ox.ac.uk/wp-content/uploads/Reframing_Superintelligence_FHI-TR-2019-1.1-1.pdf}{"Comprehensive AI Services"} model by 
Drexler, the \href{https://www.alignmentforum.org/posts/pKSmEkSQJsCSTK6nH/an-open-agency-architecture-for-safe-transformative-ai}{"Open Agency Architecture"} of Dalrymple and others. While such systems might require more engineering effort than monolithic 
neural networks trained with massive computation, that's precisely where 
computation limits help -- by making the safer, more transparent path also the more practical one.}

\paragraph{Meaningful and guaranteed human control} 

What does "strictly controllable" mean? A key idea of the "Tool" framework is to allow systems -- even if quite general and powerful -- that are guaranteed to be under meaningful human control. What does this mean? It entails two aspects. First is a design consideration: humans should be deeply and centrally involved in what the system is doing, {\em without} delegating key important decisions to the AI. This is the character of most current AI systems. Second, to the degree that AI systems are autonomous, they must have guarantees that limit their scope of action. A guarantee should be a {\em number} characterizing the probability of something happening, and a reason to believe that number. This is what we demand in other safety critical fields, where numbers like "mean time between failure"s and expected numbers of accidents are computed, supported, and published in safety cases.\sidenote{On safety cases in general see \href{https://onlinelibrary.wiley.com/doi/10.1002/9781119443070.ch16}{this handbook}. Pertaining to AI in particular, see \href{https://papers.ssrn.com/sol3/papers.cfm?abstract_id=4806274}{Wasil et al.}, \href{https://arxiv.org/abs/2403.10462}{Clymer et al.}, \href{https://arxiv.org/abs/2410.21572}{Buhl et al.}, and \href{https://arxiv.org/abs/2411.03336}{Balesni et al.}}
The ideal number for failures is zero, of course. And the good news is that we might get quite close, albeit using quite different AI architectures, using ideas of {\em formally verified} properties of programs (including AI). The idea, explored at length by Omohundro, Tegmark, Bengio, Dalrymple, and others (see \href{https://arxiv.org/abs/2309.01933}{here} and \href{https://arxiv.org/abs/2405.06624}{here})
is to construct a program with certain properties (for example: that a human can shut it down) and formally {\em prove} that those properties hold. This can be done now for quite short programs and simple properties, but the (coming) power of AI-powered proof software could allow it for much more complex programs (e.g. wrappers) and even AI itself. 
This is a very ambitious program, but as pressure grows on the Gates, we're going to need some powerful materials reinforcing them. Mathematical proof may be one of the few that is strong enough.

\paragraph{Wither the AI industry}

With AI progress redirected, Tool AI would still be an enormous industry. In terms of hardware, even with compute caps to prevent superintelligence, training and inference in smaller models will still require huge amounts of specialized components. On the software side, defusing the explosion in AI model and computation size should simply lead to companies redirecting resources toward making the smaller systems better, more diverse, and more specialized, rather than simply making them bigger.\sidenote[][-35mm]{We are in fact already seeing this trend driven just by the high cost of inference: smaller and more specialized models "distilled" from larger ones and capable of running on less expensive hardware.}  There would be plenty of room -- more probably -- for all those money-making Silicon Valley startups.\sidenote[][-20mm]{I understand why those excited about the AI tech ecosystem may oppose what they see as onerous regulation on their industry. But it is frankly baffling to me why, say, a venture capitalist would want to allow runaway to AGI and superintelligence.  Those systems (and companies, while they remain under company control) will {\em eat all of the startups as a snack}. Probably even {\em sooner} than eating other industries. Anyone invested in a thriving AI ecosystem should prioritize ensuring that AGI development does not lead to monopolization by a few dominant players.}

\subsection*{Tool AI can yield (almost) everything humanity wants, without AGI}

Intelligence, whether biological or machine, can be broadly considered as the ability to plan and execute activities bringing about futures more in line with a set of goals.  As such, intelligence is of enormous benefit when used in pursuit of wisely chosen goals. Artificial intelligence is attracting huge investments of time and effort largely because of its promised benefits. So we should ask: to what degree would we still garner the benefits of AI if we contain its runaway to superintellience? 
The answer: we may lose surprisingly little. 

Consider first that current AI systems are already very powerful, and we have really only scratched the surface of what can be done with them.\sidenote[][-28mm]{As economist and former Deepmind researcher Michael Webb \href{https://80000hours.org/podcast/episodes/michael-webb-ai-jobs-labour-market/}{put it},
"I think if we stopped all development of bigger language models today, so GPT-4 and Claude and whatever, and they’re the last things that we train of that size -- so we’re allowing lots more iteration on things of that size and all kinds of fine-tuning, but nothing bigger than that, no bigger advancements -- just what we have today I think is enough to power 20 or 30 years of incredible economic growth."} They are reasonably capable of "running the show" in terms of "understanding" a question or task presented to them, and what it would take to answer this question or do that task.

Next, much of the excitement about modern AI systems is due to their generality; but some of the most capable AI systems -- such as ones that generate or recognize speech or images, do scientific prediction and modeling, play games, etc. -- are much narrower and well "within the Gates" in terms of computation.\sidenote[][-15mm]{For example, DeepMind's alphafold system used only 100,000th of GPT-4's FLOP number.} These systems are super-human at the particular tasks they do. They may have edge-case\sidenote[][-10mm]{The difficulty of self-driving cars is important to note here: while nominally a narrow task, and achievable with fair reliability with relatively small AI systems, extensive real-world knowledge and understanding is necessary to get reliability to the level needed in such a safety-critical task.} (or \href{https://arxiv.org/abs/2211.00241}{exploitable}) weaknesses due to their narrowness; however {\em totally} narrow or {\em fully} general are not the only options available: there are many architectures in between.\sidenote[][1mm]{For example, given a computation budget, we'd likely see GPAI models pre-trained at (say) half that budget, and the other half used to train up very high capability in a more narrow range of tasks.  This would give super-human narrow capability backstopped by near-human general intelligence.}  

These AI tools can greatly speed advancement in other positive technologies, without AGI. To do better nuclear physics, we don't need AI to be a nuclear physicist -- we have those! If we want to accelerate medicine, give the biologists, medical researchers, and chemists powerful tools. They want them and will use them to enormous gain. We don't need a server farm full of a million digital geniuses; we have millions of humans whose genius AI can help bring out. Yes, it will take longer to get immortality and the cure to all diseases. This is a real cost. But even the most promising health innovations would be of little use if AI-driven instability leads to global conflict or societal collapse. We owe it to ourselves to give AI-empowered humans a go at the problem first.

And suppose there is, in fact, some enormous upside to AGI that cannot be obtained by humanity using in-Gate tools. Do we lose those by {\em never} building AGI and superintelligence? In weighing the risks and rewards here, there is an enormous asymmetric benefit in waiting versus rushing: we can wait until it can be done in a guaranteed safe and beneficial way,
and almost everyone will still get to reap the rewards; if we rush, it could be -- in the words of the OpenAI CEO Sam Altman -- \href{https://www.businessinsider.com/chatgpt-openai-ceo-worst-case-ai-lights-out-for-all-2023-1?op=1}{lights out for {\em all} of us.}

But if non-AGI tools are potentially so powerful, can we manage {\rm them}? The answer is a clear...maybe.

\subsection*{Tool AI systems are (probably, in principle) manageable}

But it will not be easy.  Current cutting-edge AI systems can greatly empower people and institutions in achieving their goals.  This is, in general, a good thing!  However, there are natural dynamics of having such systems at our disposal -- suddenly and without much time for society to adapt -- that offer serious risks that need to be managed.  It is worth discussing a few major classes of such risks, and how they may be diminished, assuming a Gate closure.

One class of risks is of high-powered Tool AI allowing access to knowledge or capability that had previously been tied to a person or organization, making a combination of high capability plus high loyalty available to a very broad array of actors. Today, with enough money a person of ill intent could hire a team of chemists to design and produce new chemical weapons -- but it isn't so very easy to have that money or to find/assemble the team and convince them to do something pretty clearly illegal, unethical, and dangerous. To prevent AI systems from playing such a role, improvements on current methods may well suffice,\sidenote[][-300pt]{The current dominant alignment technique is "reinforcement learning by human feedback" \href{https://arxiv.org/abs/1706.03741}{(RLHF)} and uses human feedback to create a reward/punishment signal for reinforcement leaning of the AI model. This and related techniques like \href{https://arxiv.org/abs/2212.08073}{constitutional AI} are working surprisingly well (though they lack robustness and can be circumvented with modest effort.) In addition, current language models are generally competent enough at common-sense reasoning that they will not make foolish moral mistakes. This is something of a sweet spot: smart enough to understand what people want (to the degree it can be defined), but not smart enough to plan elaborate deceptions or cause huge harm when they get it wrong.} as long as all those systems and access to them are responsibly managed. On the other hand, if powerful systems are released for general use and modification, any built-in safety measures are likely removable. So to avoid risks in this class, strong restrictions as to what can be publicly released -- analogous to restrictions on details of nuclear, explosive, and other dangerous technologies -- will be required.\sidenote[][-30mm]{In the long run, any level of AI capability that gets developed is likely to proliferate, since ultimately it is software, and useful. We'll need to have robust mechanisms to defend against the risks such systems posed.  But we {\em do not have that now} so we must be very measured in how much powerful AI models are allowed to proliferate.} 

A second class of risks stems from the scaling up of machines that act like or impersonate people. At the level of harm to individual people, these risks include much more effective scams, spam, and phishing, and the proliferation of non-consensual deepfakes.\sidenote[][20pt]{The vast majority of these are non-consensual pornographic deepfakes, including of minors.} At a collective level, they include disruption of core social processes like public discussion and debate, our societal information and knowledge gathering, processing, and dissemination systems, and our political choice systems. Mitigating this risk is likely to involve (a) laws restricting the impersonation of people by AI systems, and holding liable AI developers that create systems that generate such impersonations, (b) watermarking and provenance systems that identify and classify (responsibly) generated AI content, and (c) new socio-technical epistemic systems that can create a trusted chain from data (e.g. cameras and recordings) through facts, understanding, and good world-models.\sidenote{
Many ingredients for such solutions exist, in the form of "bot-or-not" laws (in the EU AI act among other places), \href{https://c2pa.org}{industry provenance-tracking technologies}, \href{https://www.improvethenews.org}{innovative news aggregators}, prediction \href{https://metaculus.com}{aggregators} and markets, etc.} All of this is possible, and AI can help with some parts of it.

A third general risk is that to the degree some tasks are automated, the humans presently doing those tasks can have less financial value as labor. Historically, automating tasks has made things enabled by those tasks cheaper and more abundant, while sorting the people previously doing those tasks into those still involved in the automated version (generally at higher skill/pay), and those whose labor is worth less or little. On net it is difficult to predict in which sectors more versus less human labor will be required in the resulting larger but more efficient sector. In parallel, the automation dynamic tends to increase inequality and general productivity, decrease the cost of certain goods and services (via efficiency increases), and increase the cost of others (via \href{https://en.wikipedia.org/wiki/Baumol_effect}{cost disease}). For for those on the disfavored side of the inequality increase, it is deeply unclear whether the cost decrease in those certain goods and services outweighs the increase in others, and leads to overall greater well-being.  So how will this go for AI?
Because of the relative ease with which human intellectual labor can be replaced by general AI, we can expect a rapid version of this with human-competitive general-purpose AI.\sidenote[][-30mm]{The automation wave may not follow previous patterns, in that relatively {\em high}-skill tasks such as quality writing, interpreting law, or giving medical advice, may be as much or even more vulnerable to automation than lower-skill tasks.} If we close the Gate to AGI, many fewer jobs will be wholesale replaced by AI agents; but huge labor displacement is still probable over a period of years.\sidenote[][-10mm]{For careful modeling of the effect of AGI on wages, see the report \href{https://www.imf.org/en/Publications/fandd/issues/2023/12/Scenario-Planning-for-an-AGI-future-Anton-korinek}{here}, and gory details \href{https://www.dropbox.com/scl/fi/viob7f5yv13zy0ziezlcg/AGI_Scenarios.pdf?rlkey=8hxq9rm82kksocw1zjilcxf8v&e=1&dl=0}{here}, from Anton Korinek and collaborators. They find that as more pieces of jobs are automated, productivity and wages go up -- to a point.  Once {\em too} much is automated, productivity continue to increase, but wages crater because people are replaced wholesale by efficient AI. This is why closing the Gates is so useful: we get the productivity without the vanished  human wages.} To avoid widespread economic suffering, it will likely be necessary to implement both some form of universal basic assets or income, and also engineer a cultural shift toward valuing and rewarding human-centric labor that is harder to automate (rather than seeing labor prices to drop due to the rise in available labor pushed out of other parts of the economy.) Other constructs, such as that of \href{https://hbr.org/2018/09/a-blueprint-for-a-better-digital-society}{"data dignity"} (in which the human producers of training data are auto-accorded royalties for the value created by that data in AI) may help. Automation by AI also has a second potential adverse effect, which is of {\em inappropriate} automation.  Along with applications where AI simply does a worse job, this would include those where AI systems are likely to violate moral, ethical, or legal precepts -- for example in life and death decisions, and in judicial matters.  These must be treated by applying and extending our current legal frameworks. 

Finally, a significant threat of in-gate AI is its use in personalized persuasion, attention capture, and manipulation.  We have seen in social media and other online platforms the growth of a deeply entrenched attention economy (where online services battle fiercely for user attention) and \href{https://en.wikipedia.org/wiki/The_Age_of_Surveillance_Capitalism}{"surveillance capitalism"} systems (in which user information and profiling is added to the commodification of attention.) It is all but certain that more AI will be put into the service of both.  AI is already heavily used in addictive feed algorithms, but this will evolve into addictive AI-generated content, customized to be compulsively consumed by a single person. And that person's input, responses, and data, will be fed into the attention/advertising machine to continue the vicious cycle. As well, as AI helpers provided by tech companies become the interface for more online life, they will likely replace search engines and feeds as the mechanism by which persuasion and monetization of customers occurs. Our society's failure to control these dynamics so far does not bode well. Some of this dynamic may be lessened via regulations concerning privacy, data rights, and manipulation. Getting more to the problem's root may require different perspectives, such as that of loyal AI assistants (discussed below.)

The upshot of this discussion is that of hope: in-Gate tool-based systems -- at least as long as they stay comparable in power and capability to today's most cutting-edge systems -- are probably manageable if there is will and coordination to do so. Decent human institutions, empowered by AI tools,\sidenote{There are many ways AI can be used as, and to help build, "defensive" technologies to make protections and management more robust.  See \href{https://vitalik.eth.limo/general/2025/01/05/dacc2.html}{this} influential post describing this "D/acc" agenda.} can do it. We could also fail in doing it. But it is hard to see how allowing more powerful systems would help -- other than by putting them in charge and hoping for the best.  

\vspace{-0.2cm}
\subsection*{National security}

\vspace{-0.1cm}
Races for AI supremacy -- driven by national security or other motivations -- drive us toward uncontrolled powerful AI systems that would tend to absorb, rather than bestow, power. An AGI race between the US and China is a race to determine which nation superintelligence gets first.

So what should those in charge of national security do instead?
Governments have strong experience in building controllable and secure systems, and they should double-down on doing so in AI, supporting the sort of infrastructure projects that succeed best when done at scale and with government imprimatur.

Instead of a reckless "Manhattan project" toward AGI,\sidenote[][10pt]{Somewhat ironically, a US Manhattan project would likely do little to speed timelines toward AGI -- the dial of human and fiscal investment in AI progress is already pinned at 11. The primary results would be to inspire a similar project in China (which excels at national-level infrastructure projects), to make international agreements limiting AI's risk much harder, and to alarm other geopolitical adversaries of the US such as Russia.} the US government could launch an Apollo project for controllable, secure, trustworthy systems. This could include for example:
\begin{itemize}
    \item A major program to (a) develop the on-chip hardware security mechanisms and (b) the infrastructure, to manage the compute side of powerful AI. These could build off of the US \href{https://www.commerce.gov/news/blog/2024/08/two-years-later-funding-chips-and-science-act-creating-quality-jobs-growing-local}{CHIPS act} and \href{https://www.bis.gov/press-release/biden-harris-administration-announces-regulatory-framework-responsible-diffusion}{export control regime}.
    \item A large-scale initiative to develop formal verification techniques so that particular features of AI systems (like an off-switch) can be {\em proven} to be present or absent. This can leverage AI itself to develop proofs of properties.
    \item A nation-scale effort to create software that is verifiably secure, powered by AI tools that can recode existing software into verifiably secure frameworks.
    \item A national investment project in scientific advancement using AI,\sidenote{The \href{https://nairrpilot.org/}{"National AI Research Resource"} program is a good current step in this direction and should be expanded.} running as a partnership between the DOE, NSF, and NIH.
\end{itemize}

In general, there is an enormous attack surface on our society that makes us vulnerable to risks from AI and its misuse.  Protecting from some of these risks will require government-sized investment and standardization. These would provide vastly more security than pouring gasoline on the fire of races toward AGI. And if AI is going to be built into weaponry and command-and-control systems, it is crucial that the AI be trustworthy and secure, which current AI simply is not.

\subsection*{Power concentration and its mitigations}

This essay has focused on the idea of human control of AI and its potential 
failure. But another valid lens through which to view the AI situation is 
through {\em concentration of power.} The development of very powerful AI 
threatens to concentrate power either into the very few and very large corporate 
hands that have developed and will control it, or into governments using AI as a new means to maintain their 
own power and control, or into the AI systems themselves. Or some unholy mix of the above. In any of these cases 
most of humanity loses power, control, and agency. How might we combat this?

The very first and most important step, of course, is a Gate closure to smarter-than-human AGI and superintelligence. These explicitly can directly replace humans and groups of humans.  If they are under corporate or government control they will concentrate power in those corporations or governments; if they are "free" they will concentrate power into themselves. So let's assume the Gates are closed. Then what?

One proposed 
solution to power concentration is "open-source" AI, where model weights are 
freely or widely available. But as mentioned earlier, once a model 
is open, most safety measures or guardrails can be (and generally are) stripped 
away. So there is an acute tension between on the one hand decentralization, and on the other hand safety, security, and human control of AI systems. There are also reasons to be skeptical that open models will by themselves meaningfully combat power concentration in AI any more than they have in operating systems (still dominated by Microsoft, Apple, and Google despite open alternatives).\sidenote{See \href{https://papers.ssrn.com/sol3/papers.cfm?abstract_id=4543807}{this analysis} of the various meanings and implications of "open" in tech products and how some have led to more, rather than less, entrenchment of dominance.} 

Yet there may be ways to square this circle -- to centralize and mitigate risks 
while decentralizing capability and economic reward. This requires rethinking 
both how AI is developed and how its benefits are distributed.

New models of public AI development and ownership would help. This could 
take several forms: government-developed AI (subject to democratic oversight),\sidenote{Plans in the US for a \href{https://nairratdoe.ornl.gov/}{National AI Research Resource} and the recent launch of a \href{https://fortune.com/2025/02/10/france-tech-companies-and-philanthropies-back-400-million-foundation-to-support-public-interest-ai/}{European AI Foundation} are interesting steps in this direction.} 
nonprofit AI development organizations (like Mozilla for browsers), or structures
enabling very widespread ownership and governance. Key is that these 
institutions would be explicitly chartered to serve the public interest while 
operating under strong safety constraints.\sidenote{The challenge here is not 
technical but institutional -- we urgently need real-world examples and 
experiments in what public-interest AI development could look like.} Well-crafted regulatory and standards/certifications regimes will also be vital, so that AI products offered by a vibrant market stay genuinely useful rather than exploitative toward their users. 

In terms of economic power concentration, we can use provenance tracking and "data dignity" to ensure economic benefits flow more widely. In particular, most AI power now (and in the future if we keep the Gates closed) stems from human-generated data, 
whether direct training data or human feedback. If AI companies were required to 
compensate data providers fairly,\sidenote[][-10mm]{This goes against current big tech business models and would 
require both legal action and new norms.} this could at least help distribute the economic rewards more broadly. Beyond this, another model could be public ownership of significant fractions of large AI companies. For example, governments able to tax AI companies could invest a fraction of receipts into a sovereign wealth fund that holds stock in the companies, and pays dividends to the populace.\sidenote[][-10mm]{Only some governments will be able to do so. A more radical idea is \href{https://futureoflife.org/project/the-windfall-trust/}{a universal fund of this type, under joint ownership of all humans.}}

Crucial in these mechanisms is to use the power of AI itself to help distribute power better, rather than simply fighting AI-driven power concentration using non-AI means. One powerful approach would be through well-designed AI assistants that operate with genuine fiduciary duty to their users -- putting users' interests first, especially above corporate providers'.\sidenote[][-20mm]{For a lengthy exposition of this case see \href{https://papers.ssrn.com/sol3/papers.cfm?abstract_id=3930338}{this paper} on AI loyalty. Unfortunately the default trajectory of AI assistants is likely to be one where they are increasingly disloyal.} These assistants must be truly trustworthy, technically competent yet appropriately limited based on use case and risk level, and widely available to all through public, nonprofit, or certified for-profit channels. Just as we would never accept a human assistant who secretly works against our interests for another party, we should not accept AI assistants that surveil, manipulate, or extract value from their users for corporate benefit.

Such a transformation would fundamentally alter the current dynamic where individuals are left to negotiate alone with vast (AI powered) corporate and bureaucratic machines that prioritize value extraction over human welfare. While there are many possible approaches to redistributing AI-driven power more broadly, none will emerge by default: they must be deliberately engineered and governed with mechanisms like fiduciary requirements, public provision, and tiered access based on risk.

Approaches to mitigate power concentration can face significant headwinds from incumbent powers.\sidenote{Somewhat ironically, many incumbent powers are also at risk of AI-backed disempowerment; but it may be difficult for them to perceive this until and unless the process gets quite far along.}
But there are paths toward AI development that don't require choosing between safety and concentrated power. By building the right institutions now, we could ensure that AI's benefits are widely shared while its risks are carefully managed.

\subsection*{New governance and social structures}

Our current governance structures are struggling: they are slow to respond, often 
captured by special interests, and \href{https://news.gallup.com/poll/508169/historically-low-faith-institutions-continues.aspx}{increasingly distrusted by the public.} Yet 
this is not a reason to abandon them -- quite the opposite. Some institutions 
may need replacing, but more broadly we need new mechanisms that can enhance and 
supplement our existing structures, helping them function better in our rapidly 
evolving world.

Much of our institutional weakness stems not from formal government structures, 
but from degraded social institutions: our systems for developing shared 
understanding, coordinating action, and conducting meaningful discourse. So far, 
AI has accelerated this degradation, flooding our information channels with 
generated content, pointing us to the most polarizing and divisive content, and making it harder to distinguish truth from fiction.

But AI could actually help rebuild and strengthen these social institutions. 
Consider three crucial areas:

First, AI could help restore trust in our epistemic systems -- our ways of 
knowing what is true. We could develop AI-powered systems that track and verify 
the provenance of information, from raw data through analysis to conclusions. 
These systems could combine cryptographic verification with sophisticated 
analysis to help people understand not just whether something is true, but how 
we know it's true.\sidenote{Some interesting efforts in this direction are represented by \href{https://c2pa.org/}{the c2pa coalition} on cryptographic verification; \href{https://www.improvethenews.org/}{Verity} and \href{https://ground.news/}{Ground news} on better news epistemics; and \href{metaculus.com}{Metaculus} and prediction markets on grounding discourse in falsifiable predictions.} Loyal AI assistants could be charged with following the details to ensure that they check out.

Second, AI could enable new forms of large-scale coordination. Many of our most pressing problems -- from climate change to antibiotic resistance -- are fundamentally coordination problems. We're \href{https://equilibriabook.com/}{stuck in situations that are worse than they could be for nearly everyone}, because no individual or group can afford to make the first move. AI systems could help by modeling complex incentive structures, identifying viable paths to better outcomes, and facilitating the trust-building and commitment mechanisms needed to get there.

Perhaps most intriguingly, AI could enable entirely new forms of social 
discourse. Imagine being able to "talk to a city"\sidenote{See \href{https://talktothecity.org/}{this} fascinating pilot project.} -- not just viewing 
statistics, but having a meaningful dialogue with an AI system that 
processes and synthesizes the views, experiences, needs, and aspirations of millions of 
residents. Or consider how AI could facilitate genuine dialogue between groups 
that currently talk past each other, by helping each side better understand the 
other's actual concerns and values rather than their caricatures of each other.\sidenote{See \href{https://www.kialo-edu.com/}{Kialo}, and efforts of the \href{https://www.cip.org/}{Collective Intelligence Project} for some examples.} Or AI could offer skilled, credibly neutral intermediation of disputes between people or even large groups of people (who could all interact with it directly and individually!)
Current AI is totally capable of doing this work, but the tools to do so will not come into being by themselves, or via market incentives.

These possibilities might sound utopian, especially given AI's current role in 
degrading discourse and trust. But that's precisely why we must actively develop 
these positive applications. By closing the Gates to uncontrollable AGI and prioritizing AI that enhances human agency, we can steer technological progress toward a future where AI serves as a force for empowerment, resilience, and collective advancement.

\section{The choice before us}

The last time humanity shared the Earth with other minds that spoke, thought, built technology, and did general-purpose problem solving was 40,000 years ago in ice-age Europe. Those other minds went extinct, wholly or in part due to the efforts of ours.

We are now re-entering such a time. The most advanced products of our culture and technology -- datasets built from our entire internet information commons, and 100-billion-element chips that are the most complex technologies we have ever crafted -- are being combined to bring advanced general-purpose AI systems into being.

The developers of these systems are keen to portray them as tools for human empowerment. And indeed they could be.  But make no mistake: our present trajectory is to build ever-more powerful, goal-directed, decision-making, and generally capable digital agents. They already perform as well as many humans at a broad range of intellectual tasks, are rapidly improving, and are contributing to their own improvement.

Unless this trajectory changes or hits an unexpected roadblock, we will soon -- in years, not decades -- have digital intelligences that are dangerously powerful. Even in the {\em best} of outcomes, these would bring great economic benefits (at least to some of us) but only at the cost of a profound disruption in our society, and replacement of humans in most of the most important things we do: these machines would think for us, plan for us, decide for us, and create for us. We would be spoiled, but spoiled children. Much more likely, these systems would replace humans in both the positive {\em and} negative things we do, including exploitation, manipulation, violence, and war. Can we survive AI-hypercharged versions of these?  Finally, it is more than plausible that things would not go well at all: that relatively soon we would be replaced not just in what we do, but in what we {\em are}, as architects of civilization and the future. Ask the neanderthals how that goes. Perhaps we provided them with extra trinkets for a while as well.

{\em We don't have to do this.} We have human-competitive AI, and there's no need to build AI with which we {\em can't} compete. We can build amazing AI tools without building a successor species.  he notion that AGI and superintelligence are inevitable is a
{\em choice masquerading as fate}.

By imposing some hard, global limits, we can keep AI's general capability to approximately human level while still reaping the benefits of computers' ability to process data in ways we cannot, and automate tasks none of us wants to do. These would still pose many risks, but if designed and managed well, be an enormous boon to humanity, from medicine to research to consumer products. 

Imposing limits would require international cooperation, but less than one might think, and those limits would still leave plenty of room for an enormous AI and AI hardware industry focused on applications that enhance human well-being, rather than on the raw pursuit of power. And if, with strong safety guarantees and after a meaningful global dialogue, we decide to go further, that option continues to be ours to pursue.

Humanity must {\em choose} to close the Gates to AGI and superintelligence.  

To keep the future human.

\newpage
\section{Appendixes}

\subsection{Appendix A: Compute accounting technical details}

\begin{table}[ht]
\small
\begin{tabular}{p{1.5\textwidth}}
\textbf{Compute accounting technical details}\\
\toprule
A detailed method for both "ground truth" as well as good approximations for the total compute used in training and inference is required for meaningful compute-based controls. Here is an example of how the "ground truth" could be tallied at a technical level.\\
\midrule
\textbf{Definitions:} \\
\vskip0.0in
{\em Compute causal graph:} For a given output $O$ of an AI model, there is a set of digital computations for which changing the result of that computation could potentially change $O$. (This should be conservatively assumed, i.e. there should be a clear reason to believe that a computation is independent of a precursor that both occurs earlier in time and has a physical potential causal path of effect.) This includes computation done by the AI model during inference, as well as computations that went into input, data preparation, and training of the model. Because any of these may itself be output from an AI model, this is computed recursively, cut off where a human has provided a significant change to the input.\\ \vskip0.0in
{\em Training Compute:} The total compute, in FLOP or other units, entailed by the compute causal graph of a neural network (including data preparation, training, and fine-tuning, and any other computations.) \\
\vskip0.0in
{\em Output Compute:} The total compute in the compute causal graph of a given AI output, including all neural networks (and including their Training Compute) and other computations going into that output.\\
\vskip0.0in
{\em Inference Compute Rate:} In a series of outputs, the rate of change (in FLOP/s or other units) of Output Compute between outputs, i.e. the compute used to produce the next output, divided by the timed interval between the outputs.\\
\midrule
\textbf{Examples and approximations:} \\
{\begin{itemize}
\item For a single neural network trained on human-created data, the Training Compute is simply the total training compute as customarily reported. 
\item For such a neural network doing inference at a steady rate, the Inference Compute Rate is approximately total speed of computation cluster performing the inference in FLOP/s.
\item For model fine-tuning, Training Compute of the complete model is given by the Training Compute of the non-fine-tuned model plus the computation done during fine-tuning and to prepare any data used in fine-tuning.
\item For a distilled model, the Training Compute of the complete model includes training of both the distilled model and the larger model used to provide synthetic data or other training input.
\item If several models are trained, but many "trials" are discarded on the basis of human judgment, these do not count toward the Training or Output Compute of the retained model.
\end{itemize}
}\\
\bottomrule
\end{tabular}
\label{tab:accounting}
\end{table}

\newpage
\subsection{Appendix B: Example implementation of a gate closure}


\begin{table}[ht]
\footnotesize
\begin{tabular}{p{1.5\textwidth}}
\toprule
\textbf{Implementation Example:} Here is one example of how a gate closure could work, given a limit of $10^{27}\,$ FLOP for training and $10^{20}$\,FLOP/s for inference (running the AI):\\
\midrule

\textbf{1. Pause:} For reasons of national security, the US Executive branch asks all companies based in the US, doing business in the US, or using chips manufactured in the US, to cease and desist from any new AI training runs that might exceed the $10^{27}\,$ FLOP Training Compute limit. The US should commence discussions with other countries hosting AI development, strongly encouraging them to take similar steps and indicating that the US pause may be lifted should they choose not to comply.

\textbf{2. US oversight and licensing:} By executive order or action of an existing regulatory agency, the US requires that within (say) one year:
\begin{itemize}
\item All AI training runs estimated above $10^{25}\,$ FLOP done by companies operating in the US be registered in a database maintained by a US regulatory agency. (Note: A slightly weaker version of this had already been included in the now-rescinded 2023 US executive order on AI, requiring registration for models above $10^{26}\,$ FLOP.)
\item All AI-relevant hardware manufacturers operating in the US or doing business with the USG adhere to a set of requirements on their specialized hardware and the software driving it. (Many of these requirements could be built into software and firmware updates to existing hardware, but longterm and robust solutions would require changes to later generations of hardware.) Among these is a requirement that if the hardware is part of a high-speed-interconnected cluster capable of executing $10^{18}\,$FLOP/s of computation, a higher level of verification is required, which includes regular permission by a remote "governor" who receives both telemetry and requests to perform additional computation.
\item The custodian reports the total computation performed on its hardware to the agency maintaining the US database.
\item Stronger requirements are phased in to allow both more secure and more flexible oversight and permissioning.
\end{itemize}

\textbf{3. International oversight:}
\begin{itemize}
\item The US, China, and any other countries hosting advanced chip manufacturing capability negotiate an international agreement.
\item This agreement creates a new international agency, analogous to the International Atomic Energy Agency, charged with overseeing AI training and execution.
\item Signatory countries must require their domestic AI hardware manufacturers to comply with a set of requirements at least as strong as those imposed in the US.
\item Custodians are now required to report AI computation numbers to both agencies in their home countries as well as a new office within the international agency.
\item Additional countries are strongly encouraged to join the existing international agreement: export controls by signatory counties restrict access to high-end hardware by non-signatories while signatories can receive technical support in managing their AI systems.
\end{itemize}

\textbf{4. International verification and enforcement:}
\begin{itemize}
\item The hardware verification system is updated so that it reports computation usage to both the original custodian and also directly to the international agency office.
\item The agency, via discussion with the signatories of the international agreement, agrees on computation limitations which then take legal force in the signatory countries.
\item In parallel, a set of international standards may be developed so that training and running of AIs above a threshold of computation (but below the limit) are required to adhere to those standards.
\item The agency can, if necessary to compensate for better algorithms etc., lower the computation limit. Or, if it is deemed safe and advisable (at say the level of provable safety guarantees), raise the computation limit.
\end{itemize}\\
\bottomrule
\end{tabular}
\label{tab:caps}
\end{table}

\newpage
\subsection{Appendix C: Details for a strict AGI liability regime}

\begin{table}[ht]
\small
\begin{tabular}{p{1.5\textwidth}}
\toprule
\textbf{Details for a strict AGI liability regime}\\
\midrule
{\begin{itemize}
\item {Creation and operation of an advanced AI system that is highly general, capable, and autonomous, is considered an "abnormally dangerous" activity.}
\item {As such, the default liability for training and operating such systems level is strict, joint and several liability (or its non-US equivalent) for any harms done by the model or its outputs/actions.}
\item {Personal liability will be imposed for executives and board members in cases of gross negligence or willful misconduct. This should include criminal penalties for the most egregious cases.}
\item {There are numerous safe-harbors under which liability reverts to the default (fault-based, in the US) liability to which people and companies would normally be subject.
    \begin{itemize}
    \item Models trained and operated below some compute threshold (which would be at least 10x lower than the caps described above.)
    \item AI that is "weak" (roughly, below human expert level at the tasks for which it is intended) and/or
    \item AI that is "narrow" (having a fixed and quite limited scope of tasks and operations that it is specifically designed and trained for) and/or
    \item AI that is "passive" (very limited in its ability -- even under modest modification -- to take actions or perform complex multistep tasks without direct human involvement and control.)
    \item An AI that is guaranteed to be safe, secure, and controllable (provably safe, or a risk analysis indicates a negligible level of expected harm.)
    \end{itemize}
    }
\item{Safe harbors may be claimed on the basis of a \href{https://arxiv.org/abs/2410.21572}{\em safety case} prepared by the AI developer and approved by an agency or auditor credentialed by an agency. To claim a safe harbor based on compute, the developer must just supply credible estimates of total Training Compute and maximal Inference Rate}
\item{Legislation would explicitly outline situations under which injunctive relief from the development of AI systems with a high risk of public harm would be appropriate.}
\item {Company consortia, working with NGOs and government agencies, should develop standards and norms defining these terms, how regulators should grant safe harbors, how AI developer should develop safety cases, and how courts should interpret liability where safe harbors are not proactively claimed.}
\end{itemize}
} \\
\bottomrule
\end{tabular}
\label{tab:liability}
\end{table}

\newpage
\subsection{Appendix D: A tiered approach to AGI safety \& security standards}

\begin{table}[!h]
    \small
    \renewcommand{\arraystretch}{1.2}
    \begin{tabular}{p{10mm} p{45mm} p{40mm} p{55mm}}
\toprule
\multicolumn{4}{l}{{\bf A tiered approach to AGI safety \& security standards}} \\
    \midrule
    \textbf{Risk Tier} & \textbf{Trigger(s)} & 
    \textbf{Requirements for training} & \textbf{Requirement for deployment} \\
    \midrule
    RT-0 &  AI weak in autonomy, generality, and intelligence & none & none \\
    \midrule
    RT-1 &  AI strong in one of autonomy, generality, and intelligence & none & Based on risk and use, potentially safety cases approved by national authorities wherever the model can be used \\
    \midrule
    RT-2 &  AI strong in two of autonomy, generality, and intelligence & Registration with national authority with jurisdiction over the developer & Safety case bounding risk of major harm below authorized levels plus independent safety audits (including black-box and white-box redteaming) approved by national authorities wherever the model can be used \\
    \midrule
    RT-3 & AGI strong in autonomy, generality, and intelligence & Pre-approval of safety and security plan by national authority with jurisdiction over the developer & Safety case guaranteeing bounded risk of major harm below authorized levels as well as required specifications, including cybersecurity, controllability, a non-removable killswitch, alignment with human values, and robustness to malicious use.
    \\
    \midrule
    RT-4 & Any model that also exceeds either $10^{27}$\,FLOP Training or $10^{20}$\,FLOP/s Inference & Prohibited pending international agreed lift of compute cap & Prohibited pending international agreed lift of compute cap\\
   \bottomrule

    \end{tabular}
\begin{fullwidth}

    \mycaption{\small\\ Risk classifications and safety/security standards, with tiers based on compute thresholds as well as combinations of high autonomy, generality, and intelligence:\\
    \begin{itemize}
    \item {\em Strong autonomy} applies if the system is able to perform, or can easily be made to perform, many-step tasks and/or take complex actions that are real-world relevant, without significant human oversight or intervention. Examples: autonomous vehicles and robots; financial trading bots. Non-examples: GPT-4; image classifiers\\
    \item {\em Strong generality} indicates a wide scope of application, performance of tasks for which the model was not deliberately and specifically trained, and significant ability to learn new tasks. Examples: GPT-4; mu-zero. Non-examples: AlphaFold; autonomous vehicles; image generators\\
    \item {\em Strong intelligence} corresponds to matching human expert-level performance on the tasks for which the model performs best (and for a general model, across a broad range of tasks.)  Examples: AlphaFold; mu-zero; o3. Non-examples: GPT-4; Siri \\ 
    \end{itemize}
    }
        \end{fullwidth}

    \label{tab:tiers}
\end{table}

\newpage
\section{Acknowledgments}

This work reflects the opinions of the author and should not be taken as official position of the Future of Life Institute (though they are compatible; for FLI's official position see \href{https://futureoflife.org/our-position-on-ai/#:~:text=Extreme%20Risk-,FLI%20opposes%20developing%20or%20deploying%20AI%20technologies%20that%20pose%20large,about%20incredible%20benefits%20for%20humanity.}{this page}), or any other organization with which the author is affiliated.

I'm grateful to humans Mark Brakel, Ben Eisenpress, Anna Hehir, Carlos Gutierrez, Emilia Javorsky, Richard Mallah, Jordan Scharnhorst, Elyse Fulcher, Hamza Chaudhry, Max Tegmark, Jaan Tallinn, and others for comments on the manuscript; to Tim Schrier for help with some references; to Taylor Jones and Elyse Fulcher for beautification of the manuscript and diagrams.

This work made limited use of generative AI models (Claude and ChatGPT) in its creation, for some editing and red-teaming.  In the well-established standard of levels of AI involvement of creative works, this work would probably rate a 3/10. (There is in fact no such standard! But there should be.)

\nobibliography{ktfh.bib}
\bibliographystyle{plainnat}

\end{document}